\documentclass[longnamesfirst,proof]{ectj} 

\usepackage{graphicx,grffile}
\usepackage{amsmath,amssymb}

\usepackage{amsthm}
\usepackage{mathtools,dsfont,centernot}
\usepackage{caption}
\usepackage{subcaption}
\usepackage{booktabs,tabularx,threeparttable,longtable}
\usepackage{siunitx}
\usepackage[shortlabels]{enumitem}
\usepackage{alltt}
\usepackage{hypernat}
\usepackage[nodisplayskipstretch]{setspace}
\usepackage[bottom]{footmisc}
\usepackage{xr} 
\usepackage[final]{listings}
\lstdefinestyle{inlineR}{language=R,frame=none,basicstyle=\ttfamily,keywordstyle=\ttfamily,stringstyle=\ttfamily,keepspaces=true,showspaces=false,showstringspaces=false,breaklines=true,upquote=true,print,columns=fullflexible}
\newcommand{\code}{\lstinline}

\usepackage[hyphens]{url}

\usepackage[capitalize,noabbrev]{cleveref}

\appto\TPTnoteSettings{\linespread{1}\footnotesize}

\DeclareGraphicsExtensions{.pdf,.png}
\graphicspath{ {./} {./figures/} {../figures/} }

\crefname{conjecture}{Conjecture}{Conjectures}
\crefname{section}{Section}{Sections}
\crefname{subsection}{Section}{Sections}
\crefname{subsubsection}{Section}{Sections}
\Crefname{conjecture}{Conjecture}{Conjectures}
\Crefname{section}{Section}{Sections}
\Crefname{subsection}{Section}{Sections}
\Crefname{subsubsection}{Section}{Sections}
\crefname{appendix}{Appendix}{Appendices}
\crefname{subappendix}{Appendix}{Appendices}
\crefname{subsubappendix}{Appendix}{Appendices}
\Crefname{appendix}{Appendix}{Appendices}
\Crefname{subappendix}{Appendix}{Appendices}
\Crefname{subsubappendix}{Appendix}{Appendices}
\crefname{equation}{}{}
\Crefname{equation}{Equation}{Equations}
\crefname{enumi}{}{}
\Crefname{enumi}{}{}
\renewcommand{\theenumi}{\roman{enumi}}

\crefname{assumption}{}{}
\crefname{assumption}{Assumption}{Assumptions}
\Crefname{assumption}{Assumption}{Assumptions}

\Crefname{method}{Method}{Methods}

\theoremstyle{plain}
\newtheorem{theorem}{Theorem}[section]

\newtheorem{proposition}{Proposition}[section]
\newtheorem{corollary}{Corollary}[section]

\newtheorem{assumption}{Assumption}[section]
\newtheorem{method}{Method}[section]

\newcommand{\ubar}[1]{\mkern3mu\underline{\mkern-3mu #1\mkern-3mu}\mkern3mu}
\newcommand{\matf}[1]{\ubar{\boldsymbol{\mathbf{#1}}}} 
\newcommand{\vecf}[1]{\boldsymbol{\mathbf{#1}}} 

\newcommand{\CXU}[1]{\hat{C}^U_{X(#1)}} 
\newcommand{\CYL}[1]{\hat{C}^L_{Y(#1)}} 
\newcommand{\CYU}[1]{\hat{C}^U_{Y(#1)}} 
\newcommand{\CXL}[1]{\hat{C}^L_{X(#1)}} 

\newcommand{\IQR}[2]{\tau_2\textrm{--}\tau_1\textrm{ IQR}}
\newcommand{\SD}[1]{\mathrel{\mathrm{SD}_{#1}}}
\newcommand{\nonSD}[1]{\mathrel{\mathit{non}\mathrm{SD}_{#1}}}
\newcommand{\SC}{\mathrel{\mathrm{SC}}}
\newcommand{\nonSC}{\mathrel{\mathit{non}\mathrm{SC}}}

\newcommand{\pconv}{\xrightarrow{p}}
\newcommand{\dconv}{\xrightarrow{d}}

\DeclareMathOperator{\E}{E} 
\let\Pr\relax \DeclareMathOperator{\Pr}{P} 

\DeclareMathOperator{\1}{\mathds{1}}
\newcommand{\Ind}[1]{\1\{#1\}}
\newcommand{\NormDist}{\mathrm{N}}

\newcommand{\independenT}[2]{\mathrel{\rlap{$#1#2$}\mkern2mu{#1#2}}}
\newcommand\independent{\protect\mathpalette{\protect\independenT}{\perp}} 
\providecommand{\abs}[1]{\left\lvert#1\right\rvert}

\let\originalleft\left
\let\originalright\right
\renewcommand{\left}{\mathopen{}\mathclose\bgroup\originalleft}
\renewcommand{\right}{\aftergroup\egroup\originalright}

\makeatletter
\renewenvironment{proof}[1][\proofname] {\par\pushQED{\qed}\normalfont\topsep6\p@\@plus6\p@\relax\trivlist\item[\hskip\labelsep\bfseries#1\@addpunct{:}]\ignorespaces}{\popQED\endtrivlist\@endpefalse}
\makeatother

\newcommand{\mockalph}[1]{}  
\allowdisplaybreaks[3]

\numberwithin{equation}{section}

\received{May 2022}
\accepted{TBD}
\volume{00}
\title[Comparing latent inequality with ordinal data]{Comparing latent inequality with ordinal data}

\author[D.~M.~Kaplan and W.~Zhao]{David~M.~Kaplan$^{\dagger}$ and
                Wei~Zhao$^{\dagger}$}

\address{$^{\dagger}$Department of Economics, University of Missouri, USA.}
\email{kaplandm@missouri.edu, weizhao@mail.missouri.edu}

\begin{document}

\maketitle

\begin{abstract}
We propose new ways to compare two latent distributions when only ordinal data are available and without imposing parametric assumptions on the underlying continuous distributions.
First, we contribute identification results.
We show how certain ordinal conditions provide evidence of between-group inequality, quantified by particular quantiles being higher in one latent distribution than in the other.
We also show how other ordinal conditions provide evidence of higher within-group inequality in one distribution than in the other, quantified by particular interquantile ranges being wider in one latent distribution than in the other.
Second, we propose an ``inner'' confidence set for the quantiles that are higher for the first latent distribution.
We also describe frequentist and Bayesian inference on features of the ordinal distributions relevant to our identification results.
Our contributions are illustrated by empirical examples with mental health and general health.


\end{abstract}

\section{Introduction}
\label{sec:intro}

Our results help compare latent distributions when only ordinal data are available.
For example, we want to learn about the latent health distribution, but individuals only report ordinal categories like ``poor'' or ``good,'' which each include a range of latent values between the corresponding thresholds.
Other ordinal examples include mental health, bond ratings, political indices, happiness, consumer confidence, and public school ratings.

We consider two types of inequality: within-group and between-group.
``Within-group inequality'' means dispersion, often quantified by interquantile ranges.
For example, to study whether income inequality in the U.S.\ has increased over time, a common dispersion measure is the 90--10 interquantile range, meaning the income distribution's $90$th percentile minus its $10$th percentile.
Interquantile ranges can similarly provide evidence of polarization in contexts like politics.
``Between-group inequality'' means whether one group is better off or worse off than another.
For example, ``racial inequality'' in health means one racial group tends to have better health than another.

We show how certain pairs of ordinal distributions provide evidence of either latent between-group inequality or differences in latent within-group inequality.
As in most ordered choice models, we assume the ordinal variable's value depends on the latent variable's value relative to a set of thresholds.
For within-group inequality: if the ordinal CDFs cross, then certain latent interquantile ranges are larger for one distribution, even if one group's thresholds are shifted by a constant from the other group's thresholds, providing some evidence of larger dispersion.
For between-group inequality: assuming each threshold for the second group is weakly below the corresponding threshold for the first group, the first group having a lower ordinal CDF implies certain quantiles are higher in its latent distribution.
This can be interpreted in terms of latent ``restricted stochastic dominance'' in the sense of \citet[Cond.\ I, p.\ 751]{Atkinson1987}.
All our results are robust to arbitrary increasing transformations of the latent random variables because of the equivariance property of quantiles, i.e., the quantile of the transformation equals the transformation of the quantile.

These positive results complement the negative results about comparing latent means from \citet{BondLang2019}.
In their well-published and already well-cited paper, they stress the near impossibility of comparing latent means in our continuous non-parametric framework, essentially because the mean does not share the equivariance property of quantiles.
\Citet{BondLang2019} provide conditions in Section II(A) that imply latent first-order stochastic dominance, but they note such conditions never hold in practice.
In the same framework, instead of negative results about comparing latent means, we show there is still much that can be compared between two latent distributions, specifically latent quantiles, latent restricted stochastic dominance, and latent interquantile ranges.

Our results can also be interpreted in terms of partial identification.
A pair of ordinal CDFs does not uniquely identify a pair of latent CDFs; there are an infinite number of latent CDF pairs (along with thresholds) in the identified set.
Still, sometimes all such latent CDF pairs possess a particular property, such as restricted stochastic dominance or certain interquantile ranges being larger.
Although we consider two CDFs that separately are not even partially identified due to the unknown thresholds, our within-group inequality approach is similar in spirit to that of \citet{Stoye2010}, who derives bounds for dispersion parameters of a single distribution based on the ``most compressed'' and ``most dispersed'' CDFs in the identified set.

Distinct from the ordinal inequality literature (e.g., as surveyed by \citet{Jenkins2019,Jenkins2020} or \citet{SilberYalonetzky2021}), we allow a continuous latent distribution without imposing a parametric model.
A continuous distribution is more realistic than a discrete distribution because it allows latent differences within the same ordinal category: one person with ``good'' health can be healthier than another, one A-rated bond can have higher credit worthiness than another, 
one ``pretty happy'' person may be happier than another, etc.
(Our results still apply to discrete or mixed latent distributions, too.)
A continuous latent distribution is not considered by most of the proposed ordinal inequality indexes or the median-preserving spread of \citet{AllisonFoster2004}, which \citet{Madden2014} calls ``the breakthrough in analyzing inequality with [ordinal] data'' (p.\ 206); often no latent interpretation is provided, or else the ``latent'' distribution merely allows ``rescaling'' by changing the cardinal value assigned to each category.
Similarly for happiness research, \citet[\S II(B)]{BondLang2019} write, ``Happiness researchers almost universally assume either that the ordered responses are measured on a discrete interval scale or that each group's latent happiness distribution is normal (i.e., ordered probit) or logistic (i.e., ordered logit)'' (p.\ 1634).
Such parametric specifications are usually difficult to even think about; for example, what is the shape of the latent happiness distribution?
Further, parametric models' results and conclusions are often sensitive to misspecification.
For example, \citet{BondLang2019} highlight several empirical happiness studies in which ``parametric results are reversed using plausible transformations'' (p.\ 1629).
Still, sometimes imposing a parametric model can allow other assumptions to be relaxed, like allowing random thresholds, which can be valuable and complement our approach.

With covariates, our results can be applied to pointwise comparisons of conditional distributions.
This complements the latent median regression model of \citet{ChenEtAl2021}.
For example, if we have three binary covariates, then we can compare the outcome distribution conditional on $(0,0,0)$ to the outcome distribution conditional on $(1,0,0)$, or more generally compare $(0,j,k)$ with $(1,j,k)$.
Even with a continuous covariate, we can nonparametrically estimate the ordinal distribution conditional on two distinct covariate values, then use our results to interpret the differences in terms of the latent distributions.
Or more simply, the continuous covariate can be discretized, which would also generally increase statistical precision.

For latent between-group inequality inference, we provide an ``inner'' confidence set for the true set of quantiles at which the first latent distribution is better than the second latent distribution.
This confidence set is computed from the data, and it is contained within the true set with at least the nominal probability asymptotically, as used in other economic settings by \citet[eqn.\ 1]{ArmstrongShen2015} and \citet[eqn.\ 6]{Kaplan2022consensus}.
This provides an analogue of a lower confidence bound.
That is, there is strong empirical evidence that all quantiles in the confidence set are indeed in the true set, and most likely the true set is even larger.

Our empirical examples show cases in which our results indicate evidence of latent inequality, even though latent means cannot be compared non-parametrically and latent medians are not informative.
We interpret estimated differences as well as both frequentist and Bayesian inference.

\Cref{sec:ID} contains identification results.
\Cref{sec:CS} describes the inner confidence set for latent quantile comparison.
\Cref{sec:emp} provides empirical illustrations.
%
\Cref{sec:app-pfs} collects proofs.
\Cref{sec:inf} describes frequentist and Bayesian inference.

Acronyms used include those for 
confidence set (CS), 
cumulative distribution function (CDF), 
first-order stochastic dominance (SD1), 
refined moment selection (RMS), 
and
single crossing (SC).

\section{Identification of latent relationships}
\label{sec:ID}

We state and discuss our assumptions in \cref{sec:ID-assumptions}, followed by within-group inequality results in \cref{sec:ID-within} and between-group inequality results in \cref{sec:ID-between}.

\subsection{Setting and assumptions}
\label{sec:ID-assumptions}

\Cref{a:ordinal} describes the formal setting and notation.
As in many ordered choice models, the ordinal random variables $X$ and $Y$ are derived from corresponding latent random variables $X^*$ and $Y^*$ using non-random thresholds.
For example, if $X^*$ is an individual's true latent health, then they report ``poor'' health (coded $X=1$, meaning the first category) if $X^*\le\gamma_1$, ``fair'' (coded $X=2$, meaning the second category) if $\gamma_1<X^*\le\gamma_2$, ``good'' ($X=3$) if $\gamma_2<X^*\le\gamma_3$, ``very good'' ($X=4$) if $\gamma_3<X^*\le\gamma_4$, and the highest category ``excellent'' ($X=5=J$) if $\gamma_4<X^*$.

\begin{assumption}
\label{a:ordinal}
The observable, ordinal random variables $X$ and $Y$ are derived from latent random variables $X^*$ and $Y^*$.
The $J$ ordinal categories are denoted $1,2,\ldots,J$.
(These are category labels, not cardinal values.)
The thresholds for $X$ are $-\infty = \gamma_0 < \gamma_1 < \cdots < \gamma_J = \infty$.
Using these, $X = j$ iff $\gamma_{j-1} < X^* \le \gamma_j$, also written $X = \sum_{j=1}^{J} j \Ind{\gamma_{j-1} < X^* \le \gamma_j}$, so the ordinal CDF is $F_X(j) = F_X^*(\gamma_j)$, where $F_X^*(\cdot)$ is the CDF of $X^*$.
The thresholds for $Y$ are $\gamma_j +\Delta_{\gamma,j}$, and similarly $Y = \sum_{j=1}^{J} j \Ind{\gamma_{j-1}+\Delta_{\gamma,j} < Y^* \le \gamma_j+\Delta_{\gamma,j}}$ and $F_Y(j) = F_Y^*(\gamma_j+\Delta_{\gamma,j})$, where $F_Y^*(\cdot)$ is the CDF of $Y^*$.
\end{assumption}

\begin{figure}[htbp]
\centering
\includegraphics{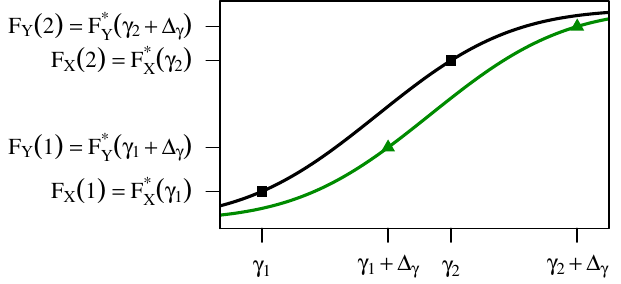}
\caption{\label{fig:basic}Example latent CDFs (lines) and ordinal CDFs (shapes).}
\end{figure}

\Cref{fig:basic} visualizes \Cref{a:ordinal}, taking $\Delta_{\gamma,1}=\Delta_{\gamma,2}$ as in the below \cref{a:thresh-shift}.
It shows how the ordinal CDF values are particular points on the latent CDFs.
Only $F_X(1)$, $F_X(2)$, $F_Y(1)$, and $F_Y(2)$ are observable; $\gamma_1$, $\gamma_2$, and $\Delta_\gamma$ are unknown.
Continuing the health example, $F_X(1)$ is the first population's proportion reporting ``poor'' health, which is done when $X^*\le\gamma_1$; $F_Y(1)$ is the second population's proportion reporting ``poor'' health, but they report ``poor'' when $Y^*\le\gamma_1+\Delta_\gamma$.

To learn about the latent distributions, there must be some restriction on the threshold differences $\Delta_{\gamma,j}$.
Otherwise, it is impossible to distinguish whether a difference in the ordinal distribution is due to a change in the latent distribution or a change in the thresholds.
For example, in \cref{fig:basic}, there is a larger proportion of $Y$ than $X$ reporting poor health, even though there is a smaller proportion of $Y^*\le\gamma_1$ than $X^*\le\gamma_1$, because of the threshold shift $\Delta_\gamma>0$.

\Cref{a:thresh-shift} allows the thresholds to differ by any amount as long as it is the same amount for each category.
For example, the $Y$ population may have higher thresholds for different health categories as long as they are all higher by the same amount.
\Cref{a:thresh-shift} is sufficient for our results on within-group inequality.

\begin{assumption}
\label{a:thresh-shift}
The thresholds for $X$ and $Y$ differ by a common constant: $\Delta_{\gamma,j}=\Delta_\gamma$, for all $j=1,\ldots,J-1$.
\end{assumption}

\Cref{a:thresh-le0} allows the thresholds to differ arbitrarily as long as each $Y$ threshold is weakly lower than the corresponding $X$ threshold.
This is sufficient for our results on between-group inequality results for evidence that $X^*$ is better than $Y^*$ because all threshold differences work in the opposite direction, making $Y$ appear better than it would if all $\Delta_{\gamma,j}=0$.

\begin{assumption}
\label{a:thresh-le0}
Each threshold for $Y$ is no greater than the corresponding threshold for $X$: $\Delta_{\gamma,j}\le0$ for all $j=1,\ldots,J-1$.
\end{assumption}

As with most assumptions, the threshold restrictions in \Cref{a:thresh-shift,a:thresh-le0} may be reasonable in some applications but not others.
For example, comparing self-reported ordinal health to the objective McMaster Health Utility Index Mark 3, \citet{LindeboomvanDoorslaer2004} find evidence of a mix of ``homogeneous reporting,'' ``index shift,'' and ``cut-point shift,'' depending on the comparison groups.
In terms of our \Cref{a:thresh-shift,a:thresh-le0}: ``homogeneous reporting'' means all $\Delta_{\gamma,j}=0$, so both \cref{a:thresh-shift,a:thresh-le0} are satisfied; ``index shift'' means there is a common non-zero $\Delta_\gamma$, so \cref{a:thresh-shift} is satisfied and \cref{a:thresh-le0} may be satisfied if $\Delta_\gamma<0$; and cut-point shift means \cref{a:thresh-shift} is violated, but possibly \cref{a:thresh-le0} holds.
Specifically, some differences across age and sex appear to violate our \cref{a:thresh-shift}, but ``for language, income and education, we find very few violations of the homogeneous reporting hypothesis, and in the few cases where it is violated, this appears almost invariably due to index rather than cut-point shift'' (p.\ 1096).
Their analysis is done conditional on a few binary variables; for example, their Table 2 tests for differences across income within the eight subgroups defined by male/female, young/old, and high/low education.
Their statistical testing assumes latent normality (pp.\ 1090--1091), but misspecification would tend to make their test more likely to falsely reject the null of homogeneous reporting or index shift.

More generally, \Cref{a:thresh-shift,a:thresh-le0} seem especially plausible when comparing the same group to itself in a different time period, as in our empirical examples in \cref{sec:emp}.
Especially if the time periods are not far apart, we may expect all $\Delta_{\gamma,j}\approx0$, so both \cref{a:thresh-shift,a:thresh-le0} are reasonable.

Further, \Cref{a:thresh-le0} is often expected when the distribution of $X$ is better than that of $Y$.
For example, if $Y$ and $X$ are older and younger adults' health, respectively, \citet{LindeboomvanDoorslaer2004} find that \cref{a:thresh-le0} holds: ``given similar objective health limitations\ldots older adults are somewhat more inclined to self-report good health'' (p.\ 1096).
That is, because older adults are generally in worse health, they lower their thresholds accordingly.
This seems natural in many situations: if $Y^*$ values tend to be lower than $X^*$, then individuals may report $Y$ based on lower thresholds than $X$.
Although this satisfies \Cref{a:thresh-le0}, such lower thresholds are still not helpful because they partially offset the latent changes.
That is, our findings are conservative in the sense of not detecting as large of a difference between $X^*$ and $Y^*$ as we would have with identical thresholds $\Delta_{\gamma,j}=0$, but we are protected against spurious findings.

\subsection{Within-group inequality}
\label{sec:ID-within}

For within-group inequality, we characterize certain pairs of ordinal CDFs that imply one latent distribution has greater dispersion than another, as quantified by latent interquantile ranges.
Throughout, we maintain \Cref{a:thresh-shift}.

\begin{figure}[htbp]
\centering
\includegraphics{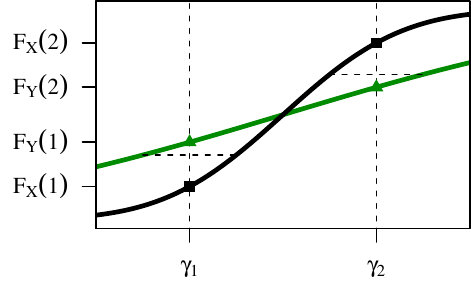}
\caption{\label{fig:SC}Illustration of \cref{res:within}.}
\end{figure}

\Cref{fig:SC} illustrates the intuition for how an ordinal CDF crossing implies a relationship between certain latent interquantile ranges.
The black line shows the latent CDF of $X^*$, with the black squares showing $F_X^*(\gamma_1)=F_X(1)$ and $F_X^*(\gamma_2)=F_X(2)$.
The green line shows the latent CDF of $Y^*$, with the green triangles showing $F_Y^*(\gamma_1)=F_Y(1)$ and $F_Y^*(\gamma_2)=F_Y(2)$, with $\Delta_\gamma=0$ for simplicity.
If instead $\Delta_\gamma\ne0$, then the difference between thresholds still remains $(\gamma_2+\Delta_\gamma)-(\gamma_1+\Delta_\gamma)=\gamma_2-\gamma_1$, so the following intuition is unchanged.

In \cref{fig:SC}, the difference between the $F_Y(2)$-quantile and $F_Y(1)$-quantile of $Y^*$ is
\begin{equation*}
Q_Y^*(F_Y(2)) - Q_Y^*(F_Y(1))
= (\gamma_2+\Delta_\gamma)-(\gamma_1+\Delta_\gamma)
= \gamma_2-\gamma_1 .
\end{equation*}
Because $F_X(1)<F_Y(1)$ and $F_X(2)>F_Y(2)$, the corresponding interquantile range for $X^*$ must be smaller.
That is, the black-with-squares line $F_X^*(\cdot)$ reaches the value $F_Y(1)$ to the right of $\gamma_1$, but reaches $F_Y(2)$ to the left of $\gamma_2$, so
\begin{equation*}
\overbrace{Q_X^*(F_Y(2))}^{<\gamma_2} - \overbrace{Q_X^*(F_Y(1))}^{>\gamma_1}
< \gamma_2 - \gamma_1
= Q_Y^*(F_Y(2)) - Q_Y^*(F_Y(1)) .
\end{equation*}

Further, consider any quantile indices $\tau_1$ and $\tau_2$ satisfying $F_X(1)<\tau_1\le F_Y(1)$ and $F_Y(2)<\tau_2\le F_X(2)$.
As seen in \cref{fig:SC}, the $Y^*$ interquantile range is larger than $\gamma_2-\gamma_1$ whereas the $X^*$ interquantile range is smaller:
\[ Q_Y^*(\tau_2)-Q_Y^*(\tau_1)
  >\gamma_2-\gamma_1
  >Q_X^*(\tau_2)-Q_X^*(\tau_1) . \]

\Cref{res:within,res:within-SC} formalize and generalize these arguments.

\begin{theorem}
\label{res:within}
Let \Cref{a:ordinal,a:thresh-shift} hold.
If there exist categories $j<k$ with $F_X(j)<F_Y(j)$ and $F_Y(k)<F_X(k)$, then $Q_X^*(\tau_2)-Q_X^*(\tau_1) < Q_Y^*(\tau_2)-Q_Y^*(\tau_1)$ for any combination of $\tau_1 \in \mathcal{T}_1=( F_X(j) , F_Y(j) ]$ and $\tau_2 \in \mathcal{T}_2=( F_Y(j) , F_X(j) ]$.
If instead $F_Y(j)<F_X(j)$ and $F_X(k)<F_Y(k)$, then $Q_Y^*(\tau_2)-Q_Y^*(\tau_1) < Q_X^*(\tau_2)-Q_X^*(\tau_1)$ instead.
\end{theorem}

\Cref{res:within} can be applied multiple times if the ordinal CDFs cross multiple times.
For example, if $F_X(1)<F_Y(1)$, $F_X(2)>F_Y(2)$, and $F_X(3)<F_Y(3)$, then it can be applied with $(j,k)$ as $(1,2)$ and $(2,3)$.
This indicates evidence of larger dispersion of $X^*$ in one part of the distribution, but larger dispersion of $Y^*$ in another.
In contrast, if the ordinal CDFs have only a single crossing, then there is evidence of only one latent distribution having larger dispersion, as in \cref{res:within-SC}.

\begin{corollary}
\label{res:within-SC}
Let \Cref{a:ordinal,a:thresh-shift} hold.
Assume a single crossing of the ordinal CDFs at category $m$ ($1\le m\le J-1$): $F_X(j) < F_Y(j)$ for $1\le j \le m$ and $F_X(j) > F_Y(j)$ for $m<j\le J-1$.
Let $Q_X^*(\cdot)$ and $Q_Y^*(\cdot)$ denote the quantile functions of $X^*$ and $Y^*$, respectively.
Then, $Q_X^*(\tau_2)-Q_X^*(\tau_1) < Q_Y^*(\tau_2)-Q_Y^*(\tau_1)$ for any combination of $\tau_2 \in \mathcal{T}_2$ and $\tau_1 \in \mathcal{T}_1$, where
\begin{equation*}
\mathcal{T}_1 \equiv \bigcup_{j=1}^{m}     ( F_X(j) , F_Y(j) ] ,
\quad
\mathcal{T}_2 \equiv \bigcup_{j=m+1}^{J-1} ( F_Y(j) , F_X(j) ]
.
\end{equation*}
\end{corollary}

\Cref{res:within-SC} interprets the median-preserving spread of \citet[pp.\ 512--513]{AllisonFoster2004} in terms of continuous latent distributions.
Analogous to a mean-preserving spread, a median-preserving spread says two ordinal distributions share the same median category but one is a ``spread out'' version of the other, i.e., can be constructed by moving probability mass away from the median.%
\footnote{There is a typo in (3c) of their definition: $X_k\ge Y_k$ should be $X_k\le Y_k$.}
Because the median-preserving spread implies a single ordinal CDF crossing, our \cref{res:within-SC} applies.
When the median differs, the median-preserving spread does not apply, but \cref{res:within-SC} still applies if there is a single ordinal CDF crossing.

Within-group inequality with ordinal data is a topic of ongoing interest.
Of the 382 Google Scholar citations of \citet{AllisonFoster2004}, over 100 have come since 2018, and their approach is only one among many for assessing inequality with ordinal data.
Recent empirical papers assessing within-group inequality from ordinal data include 
the study of durable good consumption in Asian countries by \citet[Sections 4 and 6.3]{DeutschEtAl2020}, 
``Health polarization and inequalities across Europe: an empirical approach'' by \citet{PascualEtAl2018}, 
the study of time trends in U.S.\ happiness inequality by \citet[esp.\ \S3.3]{DuttaFoster2013} and \citet{StevensonWolfers2008a}, 
and 
the cross-country comparisons of subjective well-being and education by \citet{BalestraRuiz2015}.

\subsection{Between-group inequality}
\label{sec:ID-between}

For between-group inequality (better/worse), we compare latent quantiles.
\Citet{BondLang2019} show that comparisons by latent means or latent first-order stochastic dominance are essentially impossible; quantiles provide a tractable alternative that still provide evidence of between-group inequality.
Throughout, we maintain \Cref{a:thresh-le0}.

Having a larger latent $\tau$-quantile is one piece of evidence of being ``better.''
Besides the natural intuition, this can also be interpreted in terms of quantile utility maximization (e.g., \citealp{Manski1988}; \citealp{Rostek2010}; \citealp{deCastroGalvao2019}).
For example, given strictly increasing utility function $u(\cdot)$, a $\tau$-quantile utility maximizer strictly prefers $X^*$ over $Y^*$ if the $\tau$-quantile of $u(X^*)$ is strictly greater than the $\tau$-quantile of $u(Y^*)$, which is true if and only if $Q_X^*(\tau)>Q_Y^*(\tau)$.
That is, learning $Q_X^*(\tau)>Q_Y^*(\tau)$ is equivalent to learning that $X^*$ is preferred by all $\tau$-quantile utility maximizers, regardless of utility function.
Of course, if $X^*$ is preferred for some $\tau$ but $Y^*$ is preferred for other $\tau$, then people may reasonably disagree about which is better.


The quantile intuition extends the known conclusion that latent medians can be compared if all $\Delta_{\gamma,j}=0$ and the median category differs.
That is, if the median category of ordinal $X$ is above that of $Y$, then the latent median of $X^*$ is above that of $Y^*$.
Specifically, if $\gamma_j$ is the threshold between the two categories, then the median of $Y^*$ is below $\gamma_j$ whereas the median of $X^*$ is above.
More generally, $\Delta_{\gamma,j}\le0$ is sufficient: if the $Y$ thresholds are even lower than the $X$ thresholds, then the latent median of $Y^*$ is even lower than it appears.
That is, the median of $Y^*$ is now below $\gamma_j+\Delta_{\gamma,j}$, which in turn is below $\gamma_j$, which is below the median of $X^*$.

\Cref{res:between} formally states the result for quantiles.

\begin{theorem}
\label{res:between}
Let \Cref{a:ordinal,a:thresh-le0} hold.
Let
\begin{equation*}
\begin{split}
\mathcal{T}_X &\equiv
\bigcup_{j=1}^{J-1}
\mathcal{T}_{Xj} ,\quad
\mathcal{T}_{Xj}\equiv
\left\{ \begin{array}{ll}
(F_X(j),F_Y(j)] &\textrm{if }F_X(j)<F_Y(j) ,\\
\emptyset &\textrm{otherwise.}
\end{array}
\right.
\end{split}
\end{equation*}
Then, $Q_X^*(\tau)>Q_Y^*(\tau)$ for all $\tau\in\mathcal{T}_X$.
\end{theorem}

\section{Confidence sets for latent inequality}
\label{sec:CS}

Besides using our identification results to interpret estimated ordinal distributions, we propose inner confidence sets for the latent sets defined in our results.
Specifically, interest is in the true set $\mathcal{T}_X$ in \cref{res:between}, and in the set $\mathcal{T}_1\times\mathcal{T}_2\equiv\{(\tau_1,\tau_2):\tau_1\in\mathcal{T}_1,\tau_2\in\mathcal{T}_2\}$ in \cref{res:within} and a generalization of \cref{res:within-SC}.

We first develop intuition through the special case in \cref{sec:CS-special}.
Then we describe the general between-group inequality method with formal theoretical results in \cref{sec:CS-between}.
Similarly, \cref{sec:CS-within-fixed,sec:CS-within-all} have general methods and theoretical results for within-group inequality.

An inner confidence set (CS) $\hat{\mathcal{S}}$ should be contained within the true set $\mathcal{S}$ with high asymptotic probability:
\begin{equation}
\label{eqn:inner-CS}
\Pr(\hat{\mathcal{S}}\subseteq\mathcal{S})
\ge 1-\alpha+o(1) .
\end{equation}
This general idea is used in other economic settings by \citet[eqn.\ 1]{ArmstrongShen2015} and \citet[eqn.\ 6]{Kaplan2022consensus}.

From our identification results, the sets we define are in turn subsets of the ultimate sets of interest.
Specifically, $\mathcal{T}_X$ in \cref{res:between} is a subset of $\{\tau : Q_X^*(\tau)>Q_Y^*(\tau)\}$.
Similarly, $\mathcal{T}_1\times\mathcal{T}_2$ is a subset of the set of all $(\tau_1,\tau_2)$ for which $Q_X^*(\tau_2)-Q_X^*(\tau_1)<Q_Y^*(\tau_2)-Q_Y^*(\tau_1)$.
Consequently, our inner CS for $\mathcal{T}_X$ or $\mathcal{T}_1\times\mathcal{T}_2$ is also valid for the larger full set.
In contrast, an outer CS for $\mathcal{T}_X$ or $\mathcal{T}_1\times\mathcal{T}_2$ is generally not valid for the corresponding larger set,
hence our focus on an inner CS.

\subsection{Special case: two categories}
\label{sec:CS-special}

To develop intuition, consider $J=2$, so the unknown ordinal parameters are $F_X(1)$ and $F_Y(1)$.
The true set includes all $\tau$ values above $F_X(1)$ and below $F_Y(1)$:
\begin{equation*}
\mathcal{T}_X \equiv
\left\{
\begin{array}{ll}
(F_X(1),F_Y(1)] & \textrm{if }F_X(1)<F_Y(1),\\
\emptyset & \textrm{otherwise}.
\end{array}
\right.
\end{equation*}
This set is of interest because under the conditions of \cref{res:between}, the latent $\tau$-quantile of $X^*$ is higher than that of $Y^*$ for all $\tau\in\mathcal{T}_X$.

The goal of the inner CS $\hat{\mathcal{T}}_X$ is to be contained within the true $\mathcal{T}_X$ with asymptotic probability at least $1-\alpha$: $\Pr(\hat{\mathcal{T}}_X \subseteq \mathcal{T}_X)\ge1-\alpha+o(1)$.
That is, with high probability, the true set is even larger than the inner CS.

Consider the inner CS
\begin{equation}
\label{eqn:CS-special}
\begin{split}
 \hat{\mathcal{T}}_X \equiv \left\{
\begin{array}{ll}
(\CXU{1}, \CYL{1}] & \textrm{if }\CXU{1}< \CYL{1},\\
\emptyset & \textrm{otherwise},
\end{array}
\right.
\end{split}
\end{equation}
where $ \CXU{1}$ is an upper confidence limit for $F_X(1)$, and $ \CYL{1}$ is a lower confidence limit for $F_Y(1)$, both with confidence level $\sqrt{1-\alpha}$.
That is,
\begin{equation}
\label{eqn:conf-1}
\Pr\{\CXU{1}\ge F_X(1)\}
 \ge \sqrt{1-\alpha} +o(1)
,\quad
\Pr\{\CYL{1}\le F_Y(1)\}
 \ge \sqrt{1-\alpha} +o(1).
\end{equation}
For example, if $\sqrt{n}(\hat{F}_X(1)-F_X(1)) \dconv \NormDist(0,\sigma^2)$, and $\hat\sigma^2\pconv\sigma^2$, then $\CXU{1}=\hat{F}_X(1)+z_{\sqrt{1-\alpha}}\hat\sigma/\sqrt{n}$ satisfies \cref{eqn:conf-1}, where $z_p$ denotes the $p$-quantile of the standard normal distribution.

This inner CS $\hat{\mathcal{T}}_X$ satisfies \cref{eqn:inner-CS} with asymptotic confidence level $1-\alpha$.
The argument has two steps.
First, event $\hat{\mathcal{T}}_X\subseteq \mathcal{T}_X$ is implied by the combination of events $\CXU{1}\ge F_X(1)$ and $\CYL{1}\le F_Y(1)$; that is, ``coverage'' of the inner CS is implied by coverage of both confidence limits.
Thus, if the latter combination of events occurs with at least $1-\alpha$ asymptotic probability, then so does the former event.
Second, if the two data samples are independent, then the joint coverage probability of the two confidence limits equals the product of the marginal coverage probabilities.
Formally,
\begin{align*}
&\rlap{$\overbrace{\phantom{\Pr\{\hat{\mathcal{T}}_X\subseteq \mathcal{T}_X\} \ge \Pr\{\CXU{1}\ge F_X(1), \CYL{1}\le F_Y(1) \}}}^{\textrm{because }\{\CXU{1}\ge F_X(1)\textrm{ and }\CYL{1}\le F_Y(1)\} \implies \hat{\mathcal{T}}_X\subseteq \mathcal{T}_X}$} 
\Pr\{\hat{\mathcal{T}}_X\subseteq \mathcal{T}_X\}
\ge \underbrace{\Pr\{\CXU{1}\ge F_X(1), \CYL{1}\le F_Y(1)\}= \Pr \{\CXU{1}\ge F_X(1)\}\times \Pr \{\CYL{1}\le F_Y(1)\}}_{\textrm{by independence of samples}}
\\&
 \overbrace{\ge 1-\alpha+o(1)}^{\textrm{by \cref{eqn:conf-1}}}
.
\end{align*}

\subsection{Confidence set for between-group inequality}
\label{sec:CS-between}

More generally, we want to learn about the full set $\mathcal{T}_X$ from \cref{res:between}.
Specifically, as in \cref{sec:CS-special}, we want a procedure to compute inner CS $\hat{\mathcal{T}}_X$ from data such that it is contained within the true $\mathcal{T}_X$ with asymptotic probability at least $1-\alpha$: $\Pr(\hat{\mathcal{T}}_X\subseteq\mathcal{T}_X)\ge1-\alpha+o(1)$.

Extending \cref{eqn:CS-special}, we propose the general inner CS
\begin{equation}
\label{eqn:CS-general}
\begin{split}
\hat{\mathcal{T}}_X &\equiv
\bigcup_{j=1}^{J-1}
\hat{\mathcal{T}}_{Xj},\quad \hat{\mathcal{T}}_{Xj} \equiv \left\{
\begin{array}{ll}
(\CXU{j}, \CYL{j}] & \textrm{if }\CXU{j}< \CYL{j},\\
\emptyset & \textrm{otherwise},
\end{array}
\right.
\end{split}
\end{equation}
where the $ \CXU{j}$ are joint upper confidence limits for the $F_X(j)$, and the $ \CYL{j}$ are joint lower confidence limits for the $F_Y(j)$, both with confidence level $\sqrt{1-\alpha}$.
That is,
\begin{align}
\label{eqn:CXU-general}
\Pr\{\CXU{1}\ge F_X(1), \CXU{2}\ge F_X(2), \dots, \CXU{J-1}\ge F_X(J-1)\}
&\ge \sqrt{1-\alpha}+o(1)
,\\
\label{eqn:CYL-general}
\Pr\{\CYL{1}\le F_Y(1),\CYL{2}\le F_Y(2),\dots,\CYL{J-1}\le F_Y(J-1)\}
&\ge \sqrt{1-\alpha}+o(1).
\end{align}

Theoretically, the justification follows the same two-step argument from \cref{sec:CS-special}.
First, event $\hat{\mathcal{T}}_X\subseteq \mathcal{T}_X$ is implied by all $2(J-1)$ confidence limits containing the respective true values, so the probability of the latter event is a lower bound for $\Pr(\hat{\mathcal{T}}_X\subseteq \mathcal{T}_X)$.
Second, if the two data samples are independent, then using \cref{eqn:CXU-general,eqn:CYL-general}, all $2(J-1)$ confidence limits jointly cover with probability $(\sqrt{1-\alpha}+o(1))^2=1-\alpha+o(1)$.
This is formalized in the proof of \cref{res:CS-between}.

To implement this idea, we choose the individual confidence limits to share the same pointwise coverage probability.
That is, for some $\tilde\alpha$ and all $j=1,\ldots,J-1$,
\begin{equation*}
\Pr(\CXU{j} \ge F_X(j)) = 1-\tilde\alpha+o(1)
,\quad
\Pr(\CYL{j} \ge F_X(j)) = 1-\tilde\alpha+o(1) .
\end{equation*}
In principle, the pointwise coverage probability could instead be distributed differently to reflect prior interests or beliefs, but that would both complicate the implementation and introduce opportunity for manipulation, which our implementation avoids.

Our formal results use the following assumptions.

\begin{assumption}
\label{a:ind}
The $X$ and $Y$ samples are independent: $\{X_1,\ldots,X_{n_X}\} \independent \{Y_1,\ldots,Y_{n_Y}\}$, where $n_X$ and $n_Y$ are the respective sample sizes.
\end{assumption}

\begin{assumption}
\label{a:normality}
Let $\hat{F}_X(j) \equiv n_X^{-1} \sum_{i=1}^{n_X} \Ind{X_i \le j}$, $\hat{F}_Y(j) \equiv n_Y^{-1} \sum_{i=1}^{n_Y} \Ind{Y_i \le j}$.
Letting $\hat{W}_j\equiv\hat{F}_X(j)-F_X(j)$ and $\hat{M}_j\equiv\hat{F}_Y(j)-F_Y(j)$, assume
\begin{equation*}
\sqrt{n_X} ( \hat{W}_1,\dots,\hat{W}_{J-1} )'
\dconv
\vecf{W} \sim
\NormDist( \vecf{0}, \matf{\Sigma}_X ) 
,\quad
\sqrt{n_Y} ( \hat{M}_1,\dots,\hat{M}_{J-1} )'
\dconv
\vecf{M} \sim
\NormDist( \vecf{0}, \matf{\Sigma}_Y ) ,
\end{equation*}
where the asymptotic covariance matrices have consistent estimators
\begin{equation*}
\hat{\matf{\Sigma}}_X \pconv \matf{\Sigma}_X
,\quad
\hat{\matf{\Sigma}}_Y \pconv \matf{\Sigma}_Y
.
\end{equation*}
\end{assumption}

\Cref{a:normality} is a high-level assumption that can hold across a variety of settings, including certain clustered sampling and time series settings.
To give an example of a primitive assumption, \cref{res:normality} states the sufficiency of iid sampling for \cref{a:normality}.

\begin{proposition}
\label{res:normality}
Under \cref{a:ind} and iid ordinal $X_i$ and $Y_i$, \cref{a:normality} is satisfied, with
\begin{align*}
\Sigma_{X,jk}
&\equiv
F_X(j)[1-F_X(k)] 
\textrm{ and }
\hat\Sigma_{X,jk} = \hat{F}_X(j)[1-\hat{F}_X(k)]
\textrm{ for }j \le k
,\\
\Sigma_{Y,jk}
&\equiv
F_Y(j)[1-F_Y(k)] 
\textrm{ and }
\hat\Sigma_{Y,jk} = \hat{F}_Y(j)[1-\hat{F}_Y(k)]
\textrm{ for }j \le k
,\\
\Sigma_{X,kj}&=\Sigma_{X,jk}
,\quad 
\hat\Sigma_{X,kj}=\hat\Sigma_{X,jk}
,\quad 
\Sigma_{Y,kj}=\Sigma_{Y,jk}
,\quad 
\hat\Sigma_{Y,kj}=\hat\Sigma_{Y,jk}
,
\end{align*}
where $\Sigma_{X,jk}$ is the row $j$, column $k$ element of matrix $\matf{\Sigma}_X$, $\Sigma_{Y,jk}$ is the row $j$, column $k$ element of matrix $\matf{\Sigma}_Y$, and similarly for elements of $\hat{\matf{\Sigma}}_X$ and $\hat{\matf{\Sigma}}_Y$.
\end{proposition}

\Cref{meth:CS-between} describes how to construct our inner CS.
It uses the following definitions and distributions.
Define minimum and maximum $t$-statistics as
\begin{align}
\label{eqn:def-hatTX}
\hat{T}_X \equiv \min_{j\in\{1,2,\dots,J-1\}} \{\hat{t}_X(j)\}
,\quad
\hat{t}_X(j) \equiv 
\frac{\sqrt{n_X}[\hat{F}_X(j)-F_X(j)]}{\sqrt{\hat\Sigma_{X,jj}}}
,\\
\label{eqn:def-hatTY}
\hat{T}_Y \equiv \max_{j\in\{1,2,\dots,J-1\}} \{\hat{t}_Y(j)\}
,\quad
\hat{t}_Y(j) \equiv 
\frac{\sqrt{n_Y}[\hat{F}_Y(j)-F_Y(j) ]}{\sqrt{\hat\Sigma_{Y,jj}}}
.
\end{align}
Given \Cref{a:normality}, each $\hat{t}_X(j)$ or $\hat{t}_Y(j)$ is asymptotically standard normal.
Further, the vector $(\hat{t}_X(1),\ldots,\hat{t}_X(J-1))'$ is asymptotically normal with mean zero and covariance matrix equal to the correlation matrix of $\vecf{W}$ from \cref{a:normality}; the distribution of the minimum of this normal vector is thus the asymptotic distribution of $\hat{T}_X$.
Similarly, the vector $(\hat{t}_Y(1),\ldots,\hat{t}_Y(J-1))'$ is asymptotically normal with mean zero and covariance matrix equal to the correlation matrix of $\vecf{M}$ from \cref{a:normality}, and the distribution of its maximum is the asymptotic distribution of $\hat{T}_Y$.

The following technical details show the above more formally; many readers may wish to skip them.
Define diagonal matrices $\matf{D}_X$, $\matf{D}_Y$, $\hat{\matf{D}}_X$, and $\hat{\matf{D}}_Y$ with row $j$, column $j$ entries
\begin{equation*}
\begin{split}
D_{X,jj} &= \Sigma_{X,jj}^{-1/2}
,\quad
D_{Y,jj} = \Sigma_{Y,jj}^{-1/2} 
\\
\hat{D}_{X,jj} &= \hat\Sigma_{X,jj}^{-1/2}
,\quad
\hat{D}_{Y,jj} = \hat\Sigma_{Y,jj}^{-1/2} ,
\end{split}
\end{equation*}
so
\begin{align*}
(\hat{t}_X(1),\ldots,\hat{t}_X(J-1))'
&= \hat{\matf{D}}_X \sqrt{n_X}(\hat{W}_1,\ldots,\hat{W}_{J-1})'
\dconv \matf{D}_X \vecf{W}
\sim \NormDist(\vecf{0}, \matf{D}_X\matf{\Sigma}_X\matf{D}_X)
,\\
(\hat{t}_Y(1),\ldots,\hat{t}_Y(J-1))'
&= \hat{\matf{D}}_Y \sqrt{n_Y}(\hat{M}_1,\ldots,\hat{M}_{J-1})'
\dconv \matf{D}_Y \vecf{M}
\sim \NormDist(\vecf{0}, \matf{D}_Y\matf{\Sigma}_Y\matf{D}_Y).
\end{align*}
By the continuous mapping theorem, $\hat{T}_X\dconv\min(\matf{D}_X\vecf{W})$ and $\hat{T}_Y\dconv\max(\matf{D}_Y\vecf{M})$.
Applying the continuous mapping theorem again, with $\Phi(\cdot)$ the standard normal CDF,
\begin{equation}
\label{eqn:TXhat-TYhat-asy}
\Phi(\hat{T}_X) \dconv
\Phi\bigl( \min(\matf{D}_X\vecf{W}) \bigr)
,\quad
\Phi(\hat{T}_Y) \dconv
\Phi\bigl( \max(\matf{D}_Y\vecf{M}) \bigr)
.
\end{equation}
In practice, the critical values $\tilde{\alpha}$ and $\tilde{\beta}$ in \cref{meth:CS-between} can be simulated using \cref{eqn:TXhat-TYhat-asy}, by taking many random draws from the asymptotic distributions and computing the corresponding quantiles.

\begin{method}
\label{meth:CS-between}
Construct $\hat{\mathcal{T}}_X$ as in \cref{eqn:CS-general} with
\begin{equation*}
\CXU{j}=\hat{F}_X(j) + z_{1-\tilde{\alpha}} \sqrt{\hat{\Sigma}_{X,jj}/n_X}
,\quad
\CYL{j}=\hat{F}_Y(j) - z_{1-\tilde{\beta}} \sqrt{\hat{\Sigma}_{Y,jj}/n_Y},
\end{equation*}
where $\tilde\alpha$ is the $1-\sqrt{1-\alpha}$ quantile of the asymptotic distribution of $\Phi(\hat{T}_X)$ in \cref{eqn:TXhat-TYhat-asy}, $1-\tilde\beta$ is the $\sqrt{1-\alpha}$ quantile of the asymptotic distribution of $\Phi(\hat{T}_Y)$ in \cref{eqn:TXhat-TYhat-asy}, and $\Phi(\cdot)$ is the standard normal CDF.
\end{method}

\Cref{res:CS-between} formally states the asymptotic validity of the inner CS in \cref{meth:CS-between}.

\begin{theorem}
\label{res:CS-between}
Under \cref{a:ind,a:normality} and the definition of $\mathcal{T}_X$ in \cref{res:between}, inner CS $\hat{\mathcal{T}}_X$ from \cref{meth:CS-between} satisfies $\Pr(\hat{\mathcal{T}}_X\subseteq\mathcal{T}_X)\ge 1-\alpha+o(1)$.
\end{theorem}

\subsection{Confidence set for within-group inequality: fixed category pair}
\label{sec:CS-within-fixed}

We first consider a confidence set corresponding to \cref{res:within} for a predetermined pair of categories $j<k$.
The population object of interest is the set $\mathcal{T}_{Xj} \times \mathcal{T}_{Yk}$, using notation from \cref{res:between} and similarly defining $\mathcal{T}_{Yk}$ as the interval $(F_Y(k),F_X(k)]$ if $F_Y(k)<F_X(k)$ (and the empty set otherwise).
Given the assumptions of \cref{res:within}, $Q_X^*(\tau_2)-Q_X^*(\tau_1) < Q_Y^*(\tau_2)-Q_Y^*(\tau_1)$ for all $(\tau_1,\tau_2)\in\mathcal{T}_{Xj}\times\mathcal{T}_{Yk}$.

This is almost the same as in \cref{sec:CS-between}, except the $X$ and $Y$ are switched at category $k$, so the confidence limits should be switched, too.
To be rigorous, we give the details here.
Let
\begin{equation*}
\hat{T}_{X,j,k} \equiv
\max\{ -\hat{t}_X(j), \hat{t}_X(k) \}
,\quad
\hat{T}_{Y,j,k} \equiv
\max\{ \hat{t}_Y(j), -\hat{t}_Y(k) \} ,
\end{equation*}
with the $t$-statistics defined in \cref{eqn:def-hatTX,eqn:def-hatTY}.
Analogous to \cref{eqn:TXhat-TYhat-asy}, given \Cref{a:normality},
\begin{equation}
\label{eqn:Thatjk-asy}
\begin{split}
\Phi(\hat{T}_{X,j,k})
&\dconv \Phi\bigl( \max\{ (-\matf{D}_X\vecf{W})_j , (\matf{D}_X\vecf{W})_k \} \bigr)
\\
\Phi(\hat{T}_{Y,j,k})
&\dconv \Phi\bigl( \max\{ (\matf{D}_Y\vecf{M})_j , (-\matf{D}_Y\vecf{M})_k \} \bigr) ,
\end{split}
\end{equation}
the quantiles of which can be simulated like before.

\begin{method}
\label{meth:CS-within-fixed}
Let $1-\tilde\alpha$ be the $\sqrt{1-\alpha}$ quantile of the asymptotic distribution of $\Phi(\hat{T}_{X,j,k})$ in \cref{eqn:Thatjk-asy}, and let $1-\tilde\beta$ be the $\sqrt{1-\alpha}$ quantile of the asymptotic distribution of $\Phi(\hat{T}_{Y,j,k})$ in \cref{eqn:Thatjk-asy}.
Let
\begin{align*}
\CXU{j} 
&= \hat{F}_X(j) + z_{1-\tilde{\alpha}} \sqrt{\hat{\Sigma}_{X,jj}/n_X}
,
& \CYL{j} &= \hat{F}_Y(j) - z_{1-\tilde{\beta}} \sqrt{\hat{\Sigma}_{Y,jj}/n_Y}
,\\
\CXL{k}
&= \hat{F}_X(k) - z_{1-\tilde{\alpha}} \sqrt{\hat{\Sigma}_{X,kk}/n_X}
,
& \CYU{k} &= \hat{F}_Y(k) + z_{1-\tilde{\beta}} \sqrt{\hat{\Sigma}_{Y,kk}/n_Y}
.
\end{align*}
If $\CXU{j}\ge\CYL{j}$ or $\CYU{k}\ge\CXL{k}$, then the inner CS is empty, otherwise the inner CS is
\begin{equation*}
\hat{\mathcal{T}}_{Xj}
\times
\hat{\mathcal{T}}_{Yk}
\equiv
(\CXU{j},\CYL{j}]
\times
(\CYU{k},\CXL{j}] .
\end{equation*}
\end{method}

\begin{theorem}
\label{res:CS-within-fixed}
Under \cref{a:ind,a:normality}, the inner CS in \cref{meth:CS-within-fixed} satisfies $\Pr(\hat{\mathcal{T}}_{Xj}\times\hat{\mathcal{T}}_{Yk}\subseteq\mathcal{T}_{Xj}\times\mathcal{T}_{Yk})\ge 1-\alpha+o(1)$.
\end{theorem}

\subsection{Confidence set for within-group inequality: all categories}
\label{sec:CS-within-all}

To consider all possible category pairs, the details are more complicated, but the intuition remains the same.
The strategy again is to first construct joint confidence limits for the ordinal CDFs, although this time they are two-sided.
Then, among all pairs of ordinal CDFs within the confidence limits, we find the pair that generates the smallest set of $(\tau_1,\tau_2)$ with corresponding interquantile range smaller for $X^*$ than for $Y^*$.
This set is contained within the corresponding set for any other pair within the confidence limits, which includes the true pair of ordinal CDFs with probability $1-\alpha$; thus, this set is an inner confidence set.
This approach is valid but may be conservative in some cases; it would be valuable to refine its precision in future work.


To be explicit, we construct an inner CS for the population set
\begin{align}
\label{eqn:CS-within-all-obj}
\mathcal{T} &\equiv \{ (\tau_1,\tau_2) :
  \textrm{for some }j<k,
  \tau_1\in\mathcal{T}_{Xj}\textrm{ and }%
  \tau_2\in\mathcal{T}_{Yk} \}
,
\end{align}
where for each $j=1,\ldots,J-1$,
\begin{align*}
\mathcal{T}_{Xj} &\equiv \left\{
\begin{array}{ll}
(F_X(j), F_Y(j)] & \textrm{if }F_X(j)< F_Y(j),\\
\emptyset & \textrm{otherwise},
\end{array}
\right.
\\
\mathcal{T}_{Yj} &\equiv \left\{
\begin{array}{ll}
(F_Y(j), F_X(j)] & \textrm{if }F_Y(j)< F_X(j),\\
\emptyset & \textrm{otherwise}.
\end{array}
\right.
\end{align*}
If there is a single crossing, then this $\mathcal{T}$ matches \cref{res:within-SC}'s $\mathcal{T}_1\times\mathcal{T}_2$, but this $\mathcal{T}$ is well-defined even without a single crossing.
Like before, the goal is to construct an inner CS $\hat{\mathcal{T}}$ such that $\Pr(\hat{\mathcal{T}}\subseteq\mathcal{T})\ge1-\alpha+o(1)$.

To derive two-sided confidence limits, consider the maximum absolute $t$-statistics,
\begin{equation*}
\hat{T}_{\abs{X}} \equiv
\max_{j\in\{1,\ldots,J-1\}}
\abs{\hat{t}_X(j)}
,\quad
\hat{T}_{\abs{Y}} \equiv
\max_{j\in\{1,\ldots,J-1\}}
\abs{\hat{t}_Y(j)} ,
\end{equation*}
where $\hat{t}_X(j)$ and $\hat{t}_Y(j)$ are from \cref{eqn:def-hatTX,eqn:def-hatTY}.
Analogous to \cref{eqn:TXhat-TYhat-asy}, given \Cref{a:normality},
\begin{equation}
\label{eqn:That-abs-asy}
\begin{split}
\Phi(\hat{T}_{\abs{X}})
&\dconv \Phi\bigl( \max(\abs{\matf{D}_X\vecf{W}}) \bigr)
,\quad
\Phi(\hat{T}_{\abs{Y}})
 \dconv \Phi\bigl( \max(\abs{\matf{D}_Y\vecf{M}}) \bigr) ,
\end{split}
\end{equation}
the quantiles of which can be simulated like before.

\begin{method}
\label{meth:CS-within-all}
Let $1-\tilde\alpha/2$ be the $\sqrt{1-\alpha}$ quantile of the asymptotic distribution of $\Phi(\hat{T}_{\abs{X}})$ in \cref{eqn:That-abs-asy}, and let $1-\tilde\beta/2$ be the $\sqrt{1-\alpha}$ quantile of the asymptotic distribution of $\Phi(\hat{T}_{\abs{Y}})$ in \cref{eqn:That-abs-asy}.
For all $j=1,\ldots,J-1$, let
\begin{align*}
\CXL{j} 
&= \hat{F}_X(j) - z_{1-\tilde\alpha/2} \sqrt{\hat{\Sigma}_{X,jj}/n_X}
,
& \CYL{j} &= \hat{F}_Y(j) - z_{1-\tilde\beta/2} \sqrt{\hat{\Sigma}_{Y,jj}/n_Y}
,\\
\CXU{j}
&= \hat{F}_X(j) + z_{1-\tilde\alpha/2} \sqrt{\hat{\Sigma}_{X,jj}/n_X}
,
& \CYU{j} &= \hat{F}_Y(j) + z_{1-\tilde\beta/2} \sqrt{\hat{\Sigma}_{Y,jj}/n_Y}
.
\end{align*}
Similar to \cref{eqn:CS-general}, for all $j=1,\ldots,J-1$, let
\begin{align*}
\hat{\mathcal{T}}_{Xj} &\equiv \left\{
\begin{array}{ll}
(\CXU{j}, \CYL{j}] & \textrm{if }\CXU{j}< \CYL{j},\\
\emptyset & \textrm{otherwise},
\end{array}
\right.
\\
\hat{\mathcal{T}}_{Yj} &\equiv \left\{
\begin{array}{ll}
(\CYU{j}, \CXL{j}] & \textrm{if }\CYU{j}< \CXL{j},\\
\emptyset & \textrm{otherwise}.
\end{array}
\right.
\end{align*}
An inner CS for $\mathcal{T}$ in \cref{eqn:CS-within-all-obj} is
\begin{align*}
\hat{\mathcal{T}} \equiv
\{ (\tau_1,\tau_2) :
  \tau_1\in\hat{\mathcal{T}}_{Xj},
  \tau_2\in\hat{\mathcal{T}}_{Yk},
  j<k \} .
\end{align*}
\end{method}


\begin{theorem}
\label{res:CS-within-all}
Under \cref{a:ind,a:normality} and the definition of $\mathcal{T}$ in \cref{eqn:CS-within-all-obj}, inner CS $\hat{\mathcal{T}}$ from \cref{meth:CS-within-all} satisfies $\Pr(\hat{\mathcal{T}}\subseteq\mathcal{T})\ge 1-\alpha+o(1)$.
\end{theorem}

\section{Empirical illustrations}
\label{sec:emp}

The following empirical examples illustrate our theoretical results, showing how our results help interpret ordinal data in terms of latent relationships.
All empirical results can be replicated with the files provided on the first author's website.%
\footnote{\url{https://kaplandm.github.io}}
Code is in R (\citealp{R.core}), with help from package \code{quadprog} \Citep{R.quadprog} for our RMS implementation.
For between-group inequality, we assume $\Delta_\gamma=0$ throughout.

For estimation, we use provided sampling weights to compute appropriately weighted empirical ordinal CDFs, and then we interpret the differences in terms of latent quantiles using \cref{res:between}.

For inference, we use sampling weights when provided but (though not ideal) otherwise treat sampling as iid.
We use $\alpha=0.05$ unless otherwise noted.

For inference on ordinal first-order stochastic dominance (SD1), we use both frequentist and Bayesian methods.
The Bayesian results use the posterior probability of ordinal SD1 from a Dirichlet--multinomial model with uniform prior.
For the null of SD1, we use the refined moment selection (RMS) test of \citet{AndrewsBarwick2012} as described in \cref{sec:inf-freq-SD1}.
For the null of non-SD1, we use the intersection--union test as described in \cref{sec:inf-freq-nonSD1}.
Given some level like $\alpha=0.05$, we say there is ``statistically significant'' evidence in favor of $X\SD{1}Y$ if the posterior probability is above $1-\alpha$ (Bayesian) or the intersection--union test rejects $H_0\colon X\nonSD{1}Y$ at level $\alpha$ (frequentist), and evidence against $X\SD{1}Y$ if the posterior probability is below $\alpha$ (Bayesian) or RMS rejects $H_0\colon X\SD{1}Y$ at level $\alpha$ (frequentist).

\subsection{Data}

We study ordinal measures of mental health and general health from the popular NHIS data, available through IPUMS \Citep{IPUMSNHIS2019}.
We use the 2006 and 2008 waves because the latter is in the Great Recession while the former is not.
Besides the health variables described below, we also use the appropriate sampling weights (\code{PERWEIGHT} for studying general health; \code{SAMPWEIGHT} for mental health because it is from the supplemental survey) and the provided measures of poverty (\code{POORYN}), education (\code{EDUC}), sex (\code{SEX}), and race (\code{RACENEW}).

\subsection{Mental health}
\label{sec:emp-mental}

For mental health, we compare non-recession and recession distributions separately for men in poverty and men not in poverty.
The goal is to assess whether the Great Recession is associated with worse mental health, and whether the association is stronger for those in poverty.
Although we cannot determine the 2006 poverty status of individuals observed in 2008 because these are repeated cross-sections rather than panel data, the proportion in poverty is very similar in both years ($12.0\%$, $11.7\%$).
More potentially problematic is the significant proportions of the samples ($23\%$, $11\%$) for which the poverty measure is unavailable; we do not attempt to assess possible sample selection bias.

The mental health variable is based on the Kessler-6 scale for nonspecific psychological distress introduced by \citet{KesslerEtAl2002}, which is the sum of variables \code{AEFFORT}, \code{AHOPELESS}, \code{ANERVOUS}, \code{ARESTLESS}, \code{ASAD}, and \code{AWORTHLESS} that measure frequencies of various feelings over the past $30$ days.
We code the worst mental health as 1 and the best as 25, i.e., we subtract the raw K6 score from 25.
As a rough guide, values of 1--12 help predict serious mental illness \citep[p.\ 188]{KesslerEtAl2003}.

Our interpretations from \cref{res:between} in terms of latent quantiles are more helpful than considering the latent mean or median.
The latent means cannot be compared non-parametrically, as noted by \citet{BondLang2019}, and the median is always a very high value indicating good mental health.


\begin{figure}[htbp]
\centering
\hfill
\begin{subfigure}[b]{0.47\textwidth}
\centering
\includegraphics[width=\textwidth]{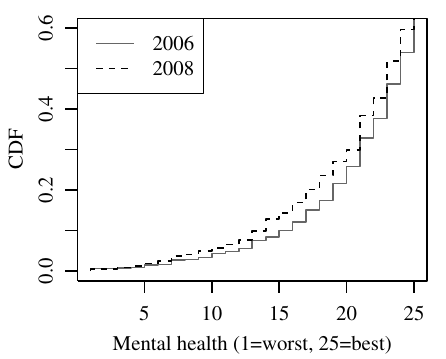}
\caption{Men in poverty}
\label{fig:K6-povT}
\end{subfigure}
\hfill
\begin{subfigure}[b]{0.47\textwidth}
\centering
\includegraphics[width=\textwidth]{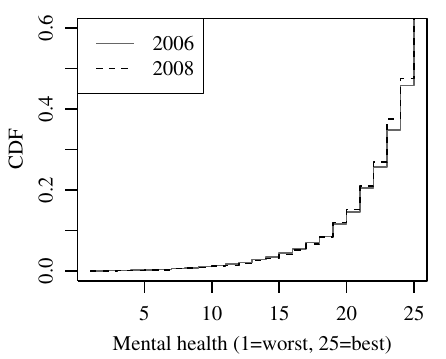}
\caption{Men not in poverty}
\label{fig:K6-povF}
\end{subfigure}
\hfill\null
\caption{\label{fig:K6}Empirical CDFs of mental health score.}
\end{figure}

\Cref{fig:K6} shows evidence of worse mental health during the Great Recession for men in poverty, which can be interpreted by applying \cref{res:between} to the estimated CDFs.
The graphs show the ordinal weighted empirical CDFs for mental health.
For men in poverty (\cref{fig:K6-povT}), the CDF for 2006 generally lies below that for 2008, indicating worse mental health in 2008 during the Great Recession.
By \cref{res:between}, the latent mental health distribution during the recession is estimated to be worse over a broad set of quantile indices $\mathcal{T}_X$ that includes $[0.03,0.30]$ as well as $[0.33,0.42]$ and other values above $0.42$ and below $0.03$.
As usual, we cannot hope to learn about changes within the lowest category, but in this case the lowest category is a small fraction of the population, well less than $1\%$.
Overall, the 2008 latent mental health distribution is estimated to be worse than 2006 over most of the lower part of the distribution.

\Cref{fig:K6} also shows the estimated 2006 and 2008 mental health ordinal CDFs for men not in poverty (\cref{fig:K6-povF}).
These appear nearly identical for most of the distribution, especially the lower half.
The only categories at which the estimated CDFs differ by at least $0.01$ are $22\le j\le 24$, and the only category where $\hat{F}_X(j)/\hat{F}_Y(j)<0.95$ is $j=23$.
That is, the changes are mostly within the part of the distribution corresponding to good mental health.

Because there are many categories ($J=25$) and the sample sizes are not very large for men in poverty ($1082$ in year 2006, $1095$ in 2008), the statistical significance is mixed.
Considering all $j=1,2,\ldots,25$, \cref{meth:CS-between} produces an empty $90\%$ inner CS.
However, restricting attention to only certain categories yields a non-empty inner CS.
For example, if the categories are combined into groups of five (1--5, 6--10, 11--15, 16--20, 21--25), then $\hat{\mathcal{T}}_X=[0.120, 0.121]$: small but not completely empty, and suggesting the strongest evidence is for a change in the lower part of the distribution.
Alternatively, if we combine categories 1--12 (which make up a small fraction of the population) and combine categories 19--25 (which correspond to pretty good mental health), but keep separate $j\in\{13,\ldots,18\}$, then the $90\%$ inner CS is
\begin{equation*}
\hat{\mathcal{T}}_X =
[0.102, 0.107] \cup
[0.119, 0.121] \cup
[0.142, 0.145] \cup
[0.173, 0.176] \cup
[0.199, 0.208] .
\end{equation*}
Again, these ranges are not large, but they reflect reasonably strong evidence of worsening mental health for men in poverty during the Great Recession, specifically in lower quantiles of the distribution.

In all, although we only have ordinal data, we can still see that the mental health costs of the Great Recession are concentrated on those in poverty, especially in lower parts of the mental health distribution.
Specifically, \cref{res:between} lets us interpret the estimated decline in mental health in terms of a broad range of the latent distribution; accounting for statistical uncertainty and being more conservative, \cref{meth:CS-between} reports a much smaller set of quantiles given a $90\%$ confidence level.

\subsection{General health}

For general health, we compare 2006 to 2008 for men in poverty, as well as comparing across racial and education groups within 2006.
The self-reported general health variable (\code{HEALTH}) has values poor, fair, good, very good, and excellent, which we code as 1, 2, 3, 4, and 5, respectively.

\begin{table}[htbp]
\centering
\caption{\label{tab:SRHS}General health comparisons.}
\sisetup{round-precision=3,round-mode=places,table-format=1.3}
\begin{threeparttable}
\begin{tabular}{cc c SSSS c c}
\toprule
Sample & Group && 
{$F(1)$} & {$F(2)$} & {$F(3)$} & {$F(4)$} && SD1?
\\
\midrule
men in poverty & 2006 && 0.0439 & 0.1560 & 0.4558 & 0.7062 && \phantom{yes}\llap{no}+\phantom{*} \\
men in poverty & 2008 && 0.0485 & 0.1656 & 0.4490 & 0.6710 && \phantom{yes}\llap{no}+\phantom{*} \\
\midrule
2006 & \phantom{high}\llap{low} edu && 0.0287 & 0.1205 & 0.3905 & 0.6731 && \phantom{yes}\llap{no}+* \\
2006 & high edu && 0.0161 & 0.0734 & 0.2957 & 0.6427 && yes+* \\
\midrule
2006 & white && 0.0212 & 0.0917 & 0.3303 & 0.6459 && yes+* \\
2006 & Black && 0.0277 & 0.1263 & 0.4150 & 0.6861 && \phantom{yes}\llap{no}+* \\
\bottomrule
\end{tabular}
\begin{tablenotes}
\item[+] Bayesian statistical significance;
          yes+ means posterior probability of SD1 (of this group over the other group) above $0.95$,
          no+ means posterior probability of SD1 below $0.05$.
\item[*] frequentist statistical significance at level $0.05$; 
          no* means RMS rejects SD1 (of this group over the other group), 
          yes* means the intersection--union test rejects non-SD1.
\end{tablenotes}
\end{threeparttable}
\end{table}

\Cref{tab:SRHS} shows the 2006/2008 comparison in the first two rows, using the sample of men in poverty as in the mental health analysis.
The first row shows the 2006 (weighted empirical) CDF evaluated at poor (1), fair (2), good (3), and very good (4); the CDF at excellent always equals $1$ by definition.
The second row shows the estimated 2008 CDF, which is higher at $1\le j\le 2$ but lower at $3\le j\le 4$.
Thus, even in the sample, there is not SD1 in either direction.

This 2006/2008 comparison of men in poverty also provides some evidence that latent health dispersion (within-group inequality) increased during the Great Recession.
In the data, the 2006 ordinal CDF crosses the 2008 ordinal CDF once from below, so \cref{res:within-SC} applies with $m=2$.
For example, the $70$--$16$ interquantile range is estimated to be larger for the latent 2008 distribution because $F_{2006}(2)<0.16<F_{2008}(2)$ and $F_{2008}(4)<0.70<F_{2006}(4)$.
\Cref{res:within-SC} holds even if $\Delta_\gamma<0$, meaning systematically lower thresholds in 2008, which could explain the larger share reporting the very best health category ($j=5$, ``excellent'').

\Cref{tab:SRHS} next compares low and high education groups with the 2006 data, with ``high'' meaning any post-secondary education.
Again, the two estimated CDFs are shown; the high education CDF is below the low education CDF at all $1\le j\le4$.
That is, the sample shows ordinal SD1, evidence of between-group inequality.
More specifically, \cref{res:between} says the latent high-education health distribution is estimated to be better at the $\tau$-quantile for (rounding to nearest $0.01$)
\[ \tau\in[0.02,0.03]\cup[0.07,0.12]\cup[0.30,0.39]\cup[0.64,0.67] .\]
Letting $X$ represent the low education ordinal health distribution and similarly $Y$ for high education, Bayesian and frequentist methods both reject $X\SD{1}Y$.
Further, there is positive evidence of $Y\SD{1}X$: the posterior probability is above $1-\alpha$, and $Y\nonSD{1}X$ is rejected by the intersection--union test at level $\alpha$ in favor of $Y\SD{1}X$.

\Cref{tab:SRHS} shows similar results for the Black/white comparison as for low/high education.
There is SD1 in the sample, 
and SD1 of Black over white ordinal health is rejected by both frequentist and Bayesian analysis, whereas SD1 of white over Black ordinal health is supported by a posterior probability above $1-\alpha$ and the intersection--union test's rejection of non-SD1 in favor of SD1.

A $90\%$ inner CS can also be computed for each comparison using \cref{meth:CS-between}.
Unlike in \cref{sec:emp-mental}, the sample sizes are very large, so the inner CS is similar to the point estimates.
For the low/high education comparison,
\begin{equation*}
\hat{\mathcal{T}}_X =
[0.018, 0.027] \cup
[0.077, 0.117] \cup
[0.302, 0.385] \cup
[0.649, 0.668] .
\end{equation*}
That is, there is strong evidence that the high-education latent health distribution is better than the low-education distribution at these quantiles.
It is probably better at other quantiles, too, but we do not have strong enough empirical evidence to say so.
For the Black/white comparison,
\begin{equation*}
\hat{\mathcal{T}}_X =
[0.023, 0.024] \cup
[0.094, 0.120] \cup
[0.335, 0.405] \cup
[0.650, 0.677] ,
\end{equation*}
indicating strong evidence that the white latent health distribution is better at these quantiles (and probably more).

\section{Conclusion}

We compare continuous latent distributions non-parametrically when only ordinal data are available.
Our identification results interpret certain ordinal patterns as evidence of between-group inequality in terms of quantiles, while other ordinal patterns indicate differences in latent within-group inequality in terms of interquantile ranges.
We propose an inner confidence set for the former set of latent quantiles.
Empirical examples with different ordinal measures of health show how our results provide insight, even when latent means cannot be compared.
Our approach can be applied similarly with ordinal measures from education, politics, finance, and other areas.
Our results can also extend to comparison of conditional distributions, which can be estimated by semiparametric or non-parametric ``distribution regression'' even with continuous regressors \citep[e.g.,][]{Frolich2006}.

\section*{Acknowledgements}

Special thanks to Longhao Zhuo for his involvement in the early stages of this project, including the contributions now published in \citet{KaplanZhuo2021} and the R code for the RMS test.
Many thanks to co-editor Petra Todd, three anonymous reviewers,
Alyssa Carlson, Denis Chetverikov, Yixiao Jiang, Jia Li, Matt Masten, Arnaud Maurel, Zack Miller, Peter Mueser, Shawn Ni, Adam Rosen, Dongchu Sun, and other participants from UConn, Duke, Yale, the 2018 Midwest Econometrics Group, and the 2019 Chinese Economists Society conference for helpful questions, comments, and references.
Thanks also to the Cowles Foundation for their hospitality during part of this work.

\section*{Conflict of Interest}

The authors have no financial interests nor any other conflicts of interest relevant to the content of this study.

\singlespacing


\appendix

\section{Proofs of Results}
\label{sec:app-pfs}

\begin{proof}[\bfseries Proof of \cref{res:within}]
For any $\tau_1\in\mathcal{T}_1$ and $\tau_2\in\mathcal{T}_2$, $F_X(j)<\tau_1\le F_Y(j)$ and $F_Y(k)<\tau_2\le F_X(k)$.
From this and \Cref{a:ordinal,a:thresh-shift},
\begin{equation*}
\begin{split}
\tau_1 &\le F_Y(j) = F_Y^*(\gamma_j+\Delta_\gamma)
\implies
Q_Y^*(\tau_1) \le \gamma_j+\Delta_\gamma ,\\
\tau_2 &> F_Y(k) = F_Y^*(\gamma_k+\Delta_\gamma)
\implies
Q_Y^*(\tau_2) > \gamma_k+\Delta_\gamma ,
\end{split}
\end{equation*}
so
\begin{equation*}
Q_Y^*(\tau_2) - Q_Y^*(\tau_1)
> (\gamma_k+\Delta_\gamma)
 -(\gamma_j+\Delta_\gamma)
= \gamma_k-\gamma_j.
\end{equation*}
Similarly,
\begin{equation*}
\begin{split}
\tau_1 &> F_X(j) = F_X^*(\gamma_j)
\implies
Q_X^*(\tau_1) > \gamma_j ,\\
\tau_2 &\le F_X(k) = F_X^*(\gamma_k)
\implies
Q_X^*(\tau_2) \le \gamma_k ,
\end{split}
\end{equation*}
so
\begin{equation*}
Q_X^*(\tau_2) - Q_X^*(\tau_1)
< \gamma_k-\gamma_j.
\end{equation*}
Altogether,
\begin{equation*}
Q_X^*(\tau_2) - Q_X^*(\tau_1)
< \gamma_k-\gamma_j
< Q_Y^*(\tau_2) - Q_Y^*(\tau_1) .
\end{equation*}

Switching the $X$ and $Y$ labels yields the converse.
\end{proof}

\begin{proof}[\bfseries Proof of \cref{res:within-SC}]
Apply \cref{res:within} to all $(j,k)$ with $j\le m<k$.
\end{proof}

\begin{proof}[\bfseries Proof of \cref{res:between}]
Consider any $\tau\in\mathcal{T}_X$, so $F_X(j)<\tau\le F_Y(j)$ for some $j$.
By \cref{a:ordinal} with $\Delta_\gamma=0$, this is equivalent to $F_X^*(\gamma_j)<\tau\le F_Y^*(\gamma_j+\Delta_{\gamma,j})$.
Because $F_Y^*(\cdot)$ is an increasing function and $\Delta_{\gamma,j}\le0$ by \Cref{a:thresh-le0}, $F_Y^*(\gamma_j+\Delta_{\gamma,j})\le F_Y^*(\gamma_j)$.
Altogether this implies $Q_Y^*(\tau)\le\gamma_j<Q_X^*(\tau)$, i.e., $Q_X^*(\tau)>Q_Y^*(\tau)$.
Switching the $X$ and $Y$ labels yields the result for $\mathcal{T}_Y$.
\end{proof}

\begin{proof}[\bfseries Proof of \cref{res:normality}]
Because $F_X(j)=\E[\Ind{X\le j}]$ and $F_Y(j)=\E[\Ind{Y\le j}]$ for all $j=1,\ldots,J-1$, the asymptotic normal distributions and covariance matrices can be derived directly with the Lindeberg--L{\'e}vy central limit theorem \citep[e.g.,][Thm.\ 6.3]{Hansen2022econometrics}.
Consistency of the covariance matrix estimators follows from the weak law of large numbers \citep[e.g.,][Thm.\ 6.1]{Hansen2022econometrics} and continuous mapping theorem \citep[e.g.,][Thm.\ 6.6]{Hansen2022econometrics}.
\end{proof}

\begin{proof}[\bfseries Proof of \cref{res:CS-between}]
First, to prove $\CXU{j}=\hat{F}_X(j) + z_{1-\tilde{\alpha}} \sqrt{\hat{\Sigma}_{X,jj}/n_X}$ from \cref{meth:CS-between} satisfies \cref{eqn:CXU-general}, the left side of \cref{eqn:CXU-general} can be rewritten as
\begin{align*}
&\Pr\{\CXU{1}\ge F_X(1), \dots,  \CXU{J-1}\ge F_X(J-1)  \}
= \Pr\{\bigcap_{j=1}^{J-1} \CXU{j}\ge F_X(j)\}
\\&\overbrace{=
   \Pr\{\bigcap_{j=1}^{J-1} \hat{F}_X(j) + \sqrt{\frac{\hat{\Sigma}_{X,jj}}{n_X}}z_{1-\tilde{\alpha}} \ge F_X(j)\}
   }^{\textrm{plug in }\CXU{j}=\hat{F}_X(j) + \sqrt{\frac{\hat{\Sigma}_{X,jj}}{n_X}}z_{1-\tilde{\alpha}}\textrm{ from \cref{meth:CS-between}}}
   =\Pr\{\bigcap_{j=1}^{J-1} \frac{\sqrt{n_X}[ \hat{F}_X(j) - F_X(j) ]}{\hat{\Sigma}_{X,jj}^{1/2}} \ge z_{\tilde{\alpha}}\}\\
&= \Pr\{ \min_{(j\in\{1,2,\dots,J-1\})} 
    \frac{\sqrt{n_X}[ \hat{F}_X(j) - F_X(j) ]}%
         {\hat{\Sigma}_{X,jj}^{1/2}}
    \ge z_{\tilde{\alpha}}\}
 \overbrace{=
   \Pr\{ \min_{(j\in\{1,2,\dots,J-1\})} \hat{t}_X(j)\ge z_{\tilde{\alpha}}\}
 }^{\textrm{notation replacement from \cref{eqn:def-hatTX}}}
\\&=
  \Pr\{\hat{T}_X \ge z_{\tilde{\alpha}}\}
= \overbrace{
  \Pr\{\Phi(\hat{T}_X) \ge \tilde{\alpha}\}
  = \sqrt{1-\alpha}+o(1)
  }^{\textrm{by definition of }\tilde\alpha}.
\label{eqn:X-joint-prob-even}\refstepcounter{equation}\tag{\theequation}
\end{align*}

Second, similarly, to prove $\CYL{j}=\hat{F}_Y(j) - z_{1-\tilde{\beta}} \sqrt{\hat{\Sigma}_{Y,jj}/n_Y}$ from \cref{meth:CS-between} satisfies \cref{eqn:CYL-general}, the left side of \cref{eqn:CYL-general} can be rewritten as
\begin{align*}
& \Pr\{\CYL{1}\le F_Y(1),\dots,\CYL{J-1}\le F_Y(J-1)\}=  \Pr\{\bigcap_{j=1}^{J-1} \CYL{j}\le F_Y(j)\}\\
&\overbrace{=\Pr\{\bigcap_{j=1}^{J-1} \hat{F}_Y(j) - \sqrt{\frac{\hat{\Sigma}_{Y,jj}}{n_Y}}z_{1-\tilde{\beta}}\le F_Y(j)\} }^{\textrm{plugged in }\CYL{j}=\hat{F}_Y(j) - \sqrt{\frac{\hat{\Sigma}_{Y,jj}}{n_Y}}z_{1-\tilde{\beta}}\textrm{ from \cref{meth:CS-between}}}=\Pr\{\bigcap_{j=1}^{J-1} \frac{\sqrt{n_Y}[ \hat{F}_Y(j)
                -F_Y(j) ]}{\hat{\Sigma}_{Y,jj}^{1/2}} \le z_{1-\tilde{\beta}}\}\\
                 &=\Pr\{ \max_{(j\in\{1,2,\dots,J-1\})}  \frac{\sqrt{n_Y}[ \hat{F}_Y(j)
                -F_Y(j) ]}{\hat{\Sigma}_{Y,jj}^{1/2}} \ge z_{1-\tilde{\beta}}\}\overbrace{=\Pr\{ \max_{(j\in\{1,2,\dots,J-1\})} \hat{t}_Y(j)\le z_{1-\tilde{\beta}}\} }^{\textrm{notation replacement from \cref{eqn:def-hatTY}}}\\
&=
\Pr\{\hat{T}_Y \le z_{1-\tilde{\beta}}\}
=
\overbrace{
  \Pr\{ \Phi(\hat{T}_Y) \le 1-\tilde{\beta} \} = \sqrt{1-\alpha}+o(1)
}^{\textrm{by definition of }1-\tilde\beta}.
\label{eqn:Y-joint-prob-even}\refstepcounter{equation}\tag{\theequation}
\end{align*}

Third, to prove $\Pr(\hat{\mathcal{T}}_X\subseteq\mathcal{T}_X)\ge 1-\alpha+o(1)$, combine the above with the fact that the joint coverage of all $2(J-1)$ confidence limits implies $\hat{\mathcal{T}}_X\subseteq \mathcal{T}_X$.
Thus, the probabilities of those events obey the inequality
\begin{align}
\label{eqn:TX-imply-general}
&\Pr\{\hat{\mathcal{T}}_X\subseteq \mathcal{T}_X\} \\
&\ge
\Pr\{\CXU{1}\ge F_X(1),\CYL{1}\le F_Y(1),\dots,  \CXU{J-1}\ge F_X(J-1),  \CYL{J-1}\le F_Y(J-1)   \}
. \notag
\end{align}
Finally,
\begin{align*}
& \overbrace{\Pr\{\hat{\mathcal{T}}_X\subseteq \mathcal{T}_X\} \ge  \Pr\{\CXU{j}\ge F_X(j),\CYL{j}\le F_Y(j)\textrm{ for all }j=1,\ldots,J-1 \}}^{\textrm{by \cref{eqn:TX-imply-general}}}\\
&\overbrace{ = \underbrace{\Pr\{\CXU{1}\ge F_X(1), \dots \CXU{J-1}\ge F_X(J-1)\}}_{=\sqrt{1-\alpha}+o(1)\textrm{ by \cref{eqn:X-joint-prob-even}}} \times \underbrace{\Pr\{\CYL{1}\le F_Y(1),\dots,\CYL{J-1}\le F_Y(J-1)\}}_{=\sqrt{1-\alpha}+o(1)\textrm{ by \cref{eqn:Y-joint-prob-even}}} }^{\textrm{under \cref{a:ind}}} \\
    &= 1-\alpha+o(1).
\end{align*}

As before, the above argument uses $\Sigma_{X,jj}>0$ and $ \Sigma_{Y,jj}>0$, for all $j\in \{1,2,\dots,J-1\}$, but the confidence limits are still valid even if $\Sigma_{X,ii}=0$ and $ \Sigma_{Y,jj}=0$ for some $i,j\in \{1,2,\dots,J-1\}$.
This is because for those $i,j\in \{1,2,\dots,J-1\}$, $\Pr (\hat{C}_{(i)} \ge F_X(i))= 1$ and $\Pr (\CYL{j} \le F_Y(j))= 1$.
Therefore, \cref{eqn:CXU-general,eqn:CYL-general} are still satisfied.
\end{proof}

\begin{proof}[\bfseries Proof of \cref{res:CS-within-fixed}]
First, we establish the joint $1-\alpha$ asymptotic confidence level of the four confidence limits.
For the $X$ limits,
\begin{align*}
& \Pr(\CXU{j}\ge F_X(j), \CXL{k}\le F_X(k))
\\&=
\Pr( \hat{F}_X(j) + z_{1-\tilde{\alpha}} \sqrt{\hat{\Sigma}_{X,jj}/n_X} \ge F_X(j) ,
\hat{F}_X(k) - z_{1-\tilde{\alpha}} \sqrt{\hat{\Sigma}_{X,kk}/n_X} \le F_X(k) )
\\&=
\Pr(z_{1-\tilde\alpha} \ge -\hat{t}_X(j) ,
    \hat{t}_X(k) \le z_{1-\tilde\alpha})
\\&=
\Pr(\max\{-\hat{t}_X(j),\hat{t}_X(k)\} \le z_{1-\tilde\alpha})
\\&=
\Pr(\Phi(\max\{-\hat{t}_X(j),\hat{t}_X(k)\}) \le 1-\tilde\alpha)
\\&=
\sqrt{1-\alpha}+o(1)
\end{align*}
by definition of $1-\tilde\alpha$, which in turn used \Cref{a:normality}.
Similarly,
\begin{align*}
& \Pr(\CYL{j}\le F_Y(j), \CYU{k}\le F_Y(k))
\\&=
\Pr( 
\hat{F}_Y(j) - z_{1-\tilde\beta} \sqrt{\hat{\Sigma}_{Y,jj}/n_Y} \le F_Y(j) ,
\hat{F}_Y(k) + z_{1-\tilde\beta} \sqrt{\hat{\Sigma}_{Y,kk}/n_Y} \ge F_Y(k)
)
\\&=
\Pr(
\hat{t}_Y(j) \le z_{1-\tilde\beta},
z_{1-\tilde\beta} \ge -\hat{t}_Y(k)
)
\\&=
\Pr(\max\{\hat{t}_Y(j), -\hat{t}_Y(k)\} \le z_{1-\tilde\beta})
\\&=
\Pr(\Phi(\max\{\hat{t}_Y(j), -\hat{t}_Y(k)\}) \le 1-\tilde\beta)
\\&=
\sqrt{1-\alpha}+o(1)
\end{align*}
Given the independence of the $X$ and $Y$ samples from \Cref{a:ind}, the joint probability is the product of the individual $\sqrt{1-\alpha}+o(1)$ probabilities, which is $1-\alpha+o(1)$:
\begin{align*}
& \Pr( \CXU{j}\ge F_X(j), \CXL{k}\le F_X(k),
    \CYL{j}\le F_Y(j), \CYU{k}\le F_Y(k) )
\\&=
\Pr( \CXU{j}\ge F_X(j), \CXL{k}\le F_X(k))
\times
\Pr( \CYL{j}\le F_Y(j), \CYU{k}\le F_Y(k) )
\\&=
(\sqrt{1-\alpha}+o(1))(\sqrt{1-\alpha}+o(1))
\\&=
1-\alpha+o(1) .
\end{align*}

Second, the final result follows because the confidence limits' joint coverage implies the inner CS is a subset of the true set.
If $\CXU{j}\ge F_X(j)$ and $\CYL{k}\le F_Y(j)$, then $\hat{\mathcal{T}}_{Xj}\subseteq\mathcal{T}_{Xj}$.
Similarly, if $\CYL{j}\le F_Y(j)$ and $\CYU{k}\ge F_Y(k)$, then $\hat{\mathcal{T}}_{Yk}\subseteq\mathcal{T}_{Yk}$.
Thus,
\begin{align*}
& \Pr(\hat{\mathcal{T}}_{Xj}\times \hat{\mathcal{T}}_{Yk}
    \subseteq \mathcal{T}_{Xj}\times\mathcal{T}_{Yk} )
\\&=
\Pr(\hat{\mathcal{T}}_{Xj} \subseteq \mathcal{T}_{Xj} ,
    \hat{\mathcal{T}}_{Yk} \subseteq \mathcal{T}_{Yk} )
\\&\ge
\Pr( \CXU{j}\ge F_X(j), \CXL{k}\le F_X(k),
    \CYL{j}\le F_Y(j), \CYU{k}\le F_Y(k) )
\\&= 1-\alpha+o(1).
\qedhere
\end{align*}
\end{proof}

\begin{proof}[\bfseries Proof of \cref{res:CS-within-all}]
The proof structure is the same as before: first, establish the asymptotic coverage probability of the joint confidence limits; second, show this provides a lower bound for the inner CS's asymptotic coverage probability.

For the $X$ confidence limits,
\begin{align*}
& \Pr( \CXL{j}\le F_X(j)\le\CXU{j}\textrm{ for all }j=1,\ldots,J-1 )
\\&=
\Pr( \abs{\hat{t}_X(j)} \le z_{1-\tilde\alpha/2}\textrm{ for all }j=1,\ldots,J-1 )
\\&=
\Pr( \hat{T}_{\abs{X}} \le z_{1-\tilde\alpha/2} )
\\&=
\Pr( \Phi(\hat{T}_{\abs{X}}) \le 1-\tilde\alpha/2 )
\\&= \sqrt{1-\alpha}+o(1) .
\end{align*}
Similarly, for the $Y$ confidence limits,
\begin{align*}
& \Pr( \CYL{j}\le F_Y(j)\le\CYU{j}\textrm{ for all }j=1,\ldots,J-1 )
\\&=
\Pr( \abs{\hat{t}_Y(j)} \le z_{1-\tilde\beta/2}\textrm{ for all }j=1,\ldots,J-1 )
\\&=
\Pr( \hat{T}_{\abs{Y}} \le z_{1-\tilde\beta/2} )
\\&=
\Pr( \Phi(\hat{T}_{\abs{Y}}) \le 1-\tilde\beta/2 )
\\&= \sqrt{1-\alpha}+o(1) .
\end{align*}
Given the independence in \Cref{a:ind}, the joint $X$ and $Y$ coverage probability is the product, $1-\alpha+o(1)$.

If all confidence limits cover their respective true ordinal CDF values, then for all $j=1,\ldots,J-1$, $\hat{\mathcal{T}}_{Xj}\subseteq\mathcal{T}_{Xj}$ and $\hat{\mathcal{T}}_{Yj}\subseteq\mathcal{T}_{Yj}$.
Thus,
\begin{align*}
& \Pr( \hat{\mathcal{T}} \subseteq \mathcal{T} )
\\&\ge
\Pr( \hat{\mathcal{T}}_{Xj}\subseteq\mathcal{T}_{Xj}\textrm{ and }\hat{\mathcal{T}}_{Yj}\subseteq\mathcal{T}_{Yj}\textrm{ for all }j=1,\ldots,J-1 )
\\&\ge
\Pr( \CXL{j}\le F_X(j)\le\CXU{j}\textrm{ and }\CYL{j}\le F_Y(j)\le\CYU{j}\textrm{ for all }j=1,\ldots,J-1 )
\\&=
\Pr( \CXL{j}\le F_X(j)\le\CXU{j}\textrm{ for all }j=1,\ldots,J-1 )
\\&\quad
\times
\Pr( \CYL{j}\le F_Y(j)\le\CYU{j}\textrm{ for all }j=1,\ldots,J-1 )
\\&=
(\sqrt{1-\alpha}+o(1))
\times
(\sqrt{1-\alpha}+o(1))
=
1-\alpha+o(1) .
\qedhere
\end{align*}
\end{proof}

\section{Statistical inference on ordinal relationships}
\label{sec:inf}

To quantify statistical uncertainty about the ordinal relationships in our identification results, we describe how to apply existing frequentist and Bayesian approaches.
We characterize the ordinal relationships as combinations of inequalities and then describe frequentist and Bayesian methods, which are then compared.

\subsection{Relationships of interest}
\label{sec:inf-H0s}

Notationally, we gather the ordinal CDF differences in
\begin{equation*}
\vecf{\theta} \equiv (\theta_1,\ldots,\theta_{J-1}) 
\textrm{ with }
\theta_j 
\equiv F_X(j)-F_Y(j) 
.
\end{equation*}

Below, we characterize the subsets of the parameter space of $\vecf{\theta}$ where various ordinal relationships hold, which involves intersections and/or unions of more basic subsets.
These characterizations help us in later sections to evaluate posterior probabilities, formulate frequentist hypothesis tests, and compare Bayesian and frequentist inference.

\subsubsection{Between-group inequality}

Ordinal first-order stochastic dominance $X \SD{1} Y$ is equivalent to:
\begin{equation}
\label{eqn:SD1}
X \SD{1} Y
\iff
\theta_j\le0\textrm{ for all }j=1,\ldots,J-1
\iff
\vecf{\theta} \in 
\bigcap_{j=1}^{J-1}
  \{ \vecf{\theta} : \theta_j\le0 \}
.
\end{equation}
The opposite of \cref{eqn:SD1} is
\begin{equation}
\label{eqn:nonSD1}
X \nonSD{1} Y
\iff
\theta_j>0\textrm{ for some }j=1,\ldots,J-1
\iff
\vecf{\theta} \in 
\bigcup_{j=1}^{J-1}
  \{ \vecf{\theta} : \theta_j>0 \}
.
\end{equation}

Ordinal SD1 has two interpretations.
Both assume $\Delta_\gamma=0$.
First, latent SD1 implies ordinal SD1, so rejecting ordinal SD1 implies rejecting latent SD1; i.e., ordinal SD1 is a testable implication of latent SD1.
Second, by \cref{res:between}, $X\SD{1}Y$ implies $\mathcal{T}_Y$ is empty because $\Ind{F_Y(j)<F_X(j)}=0$ for all $j$, whereas $\mathcal{T}_X$ may be non-empty, providing evidence of the latent $X^*$ being better than $Y^*$.

Non-SD1 is considered because from a frequentist perspective, rejecting a null hypothesis of non-SD1 in favor of SD1 is stronger evidence of SD1 than non-rejection of a null of SD1 \citep[e.g.,][p.\ 87]{DavidsonDuclos2013}.

\subsubsection{Within-group inequality}

We consider the ordinal CDF ``single crossing'' (SC) relationship from \cref{res:within-SC} that suggests the latent $Y^*$ is more dispersed than $X^*$.
SC at category $m$ ($1\le m\le J-1$) means $F_X(j)<F_Y(j)$ for $j\le m$ and $F_X(j)>F_Y(j)$ for $j>m$.
That is, $\theta_j<0$ for $j\le m$ and $\theta_j>0$ for $j>m$, which can be combined as $(2 \Ind{j\le m}-1) \theta_j < 0$ for all $j=1,\ldots,J-1$.
SC holds if these inequalities hold jointly for any value of $m$ between $1$ and $J-2$, inclusive.
With $k$ representing possible values of $m$, SC is
\begin{equation}
\label{eqn:SC}
X \SC Y
\iff
\vecf{\theta} \in
\bigcup_{k=1}^{J-2}
  \bigcap_{j=1}^{J-1} \{\vecf{\theta} : (2 \Ind{j\le k}-1) \theta_j < 0 \}
.
\end{equation}
As with non-SD1, non-SC is simply the opposite; by De Morgan's law,
\begin{equation*}
X \nonSC Y
\iff
\vecf{\theta} \in
\bigcap_{k=1}^{J-2}
  \bigcup_{j=1}^{J-1} \{\vecf{\theta} : (2 \Ind{j\le k}-1) \theta_j \ge 0 \}
.
\end{equation*}

\subsection{Frequentist hypothesis testing}
\label{sec:inf-freq}

We describe frequentist tests of the possible null hypotheses from \cref{sec:inf-H0s} under \Cref{a:ind,a:normality}.
Because the $X$ and $Y$ samples are independent (\cref{a:ind}), the corresponding CDF estimators have zero covariance.
Thus, assuming $n_X/n_Y \to \delta \in (0,\infty)$,
\begin{equation}
\label{eqn:theta-hat-dist}
\sqrt{n_X}( \hat{\vecf{\theta}} - \vecf{\theta} )
\dconv
\NormDist( \vecf{0}, \matf{\Sigma} )
, \quad
\matf{\Sigma}
\equiv \matf{\Sigma}_X + \delta \matf{\Sigma}_Y
.
\end{equation}

\subsubsection{Null hypothesis: ordinal SD1}
\label{sec:inf-freq-SD1}

Consider testing the null hypothesis $H_0\colon X\SD{1}Y$, i.e., that \cref{eqn:SD1} holds.

Although a simple Bonferroni approach is valid, recent methods like those of \citet{AndrewsBarwick2012} and \citet{RomanoShaikhWolf2014} can improve power.
Roughly speaking, instead of setting the critical value based on the least favorable configuration, they focus on the inequalities that seem ``close'' to binding in the sample.
For example, if $\theta_1$ is estimated to be very negative (e.g., $10$ standard errors below zero), then we could test only $j=2,\ldots,J-1$, which can be done with a smaller critical value and thus higher power.
\Citet[\S2.3.2]{Zhuo2017dissertation} describes how to compute the refined moment selection test of \citet{AndrewsBarwick2012} for ordinal SD1, as well as for median-preserving spread treating the median as known.
The newer methodology of \citet{CoxShi2022} also seems promising for improving power for ordinal SD1 testing.

\subsubsection{Null hypothesis: non-SD1}
\label{sec:inf-freq-nonSD1}

Consider testing $H_0\colon X\nonSD{1}Y$, i.e., that \cref{eqn:nonSD1} holds, so $X\SD{1}Y$ is now the alternative.

The intersection--union test of $H_0\colon X\nonSD{1}Y$ rejects when $H_{0j} \colon \theta_j > 0$ is rejected for all $j$.
That is, the overall rejection region is the intersection of the rejection regions for each $H_{0j}$.
If each $H_{0j}$ test has size $\alpha$, then the overall test has size $\alpha$.
See Theorem 8.3.23 and Sections 8.2.3 and 8.3.3 in \citet{CasellaBerger2002}, who also remark, ``The IUT may be very conservative'' (p.\ 306).

\subsubsection{Null hypothesis: single crossing}
\label{sec:inf-freq-SC}

Consider testing the null hypothesis $H_0\colon X\SC Y$.
Rewriting \cref{eqn:SC},
\begin{equation}
\label{eqn:H0-SC}
H_0 \colon \vecf{\theta} \in \Theta_0 , \quad
\Theta_0 \equiv \bigcup_{k=1}^{J-2} \Theta_k, \quad
\Theta_k \equiv \{ \vecf{\theta} : (2\Ind{j\le k}-1)\theta_j < 0, \, j=1,\ldots,J-1 \}
.
\end{equation}

This $H_0$ can be tested by combining the intersection--union approach of \cref{sec:inf-freq-nonSD1} with a method from \cref{sec:inf-freq-SD1}.
Each $H_{0k}\colon\vecf{\theta}\in\Theta_k$ is equivalent to $H_{0k}\colon\matf{D}\vecf{\theta}<0$ for diagonal matrix $\matf{D}$ with elements $D_{jj}=1$ for $j=1,\ldots,k$ and $D_{jj}=-1$ for $j=k+1,\ldots,J-1$ (and $D_{jk}=0$ if $j\ne k$).
Assuming $\sqrt{n_X}(\hat{\vecf{\theta}}-\vecf{\theta}) \dconv \NormDist(\vecf{0},\matf{\Sigma})$ as in \cref{eqn:theta-hat-dist} and applying the continuous mapping theorem,
\begin{equation*}
\sqrt{n_X}(\matf{D}\hat{\vecf{\theta}}-\matf{D}\vecf{\theta})
= \matf{D} \sqrt{n_X}(\hat{\vecf{\theta}}-\vecf{\theta})
\dconv \matf{D} \NormDist(\vecf{0},\matf{\Sigma})
= \NormDist(\vecf{0},\matf{D} \matf{\Sigma} \matf{D}') ,
\end{equation*}
so $\matf{D}\hat{\vecf{\theta}}$ can be used to test $\matf{D}\vecf{\theta}<\vecf{0}$ with the methods referenced in \cref{sec:inf-freq-SD1}.
That is, to test $H_0 \colon X \SC Y$ at level $\alpha$, there are two steps:%
\footnote{This is essentially the same approach suggested in Remark 5.3 of \citet{MachadoShaikhVytlacil2019}.}
\begin{enumerate}[label={\textup{\arabic*.}}, ref=\arabic*]
 \item For $k=1,\ldots,J-2$, using \cref{eqn:H0-SC}, test $H_{0k} \colon \vecf{\theta} \in \Theta_k$ at level $\alpha$ using a method from \cref{sec:inf-freq-SD1}.
 \item Reject $H_0 \colon X \SC Y$ if and only if all $H_{0k}$ are rejected.
\end{enumerate}

Frequentist tests for the related null of median-preserving spread \citep{AllisonFoster2004} are proposed and compared by \citet{AbulNagaEtAl2021}.

\subsection{Bayesian inference and frequentist comparison}
\label{sec:inf-Bayes}

Bayesian inference is relatively straightforward, given draws from the posterior distribution of the two ordinal CDFs: the proportion of draws in which a particular relationship holds is the (approximate) posterior probability of that relationship.
\Citet[\S2.3.3]{Zhuo2017dissertation} describes Bayesian inference with the Dirichlet--multinomial model (for independent, iid sampling) using the uniform prior, for both ordinal SD1 and the median-preserving spread.
\Citet{GunawanEtAl2018} provide similar results but with the improper prior as in their (7).
With non-iid sampling, other Bayesian approaches can be used, like the non-parametric Bayesian approach to complex sampling design from \citet{DongEtAl2014}.

Although frequentist confidence sets and Bayesian credible sets for $\vecf{\theta}$ tend to agree in large samples, frequentist and Bayesian assessments of SD1 and SC can differ.
Writing the null hypothesis as $H_0\colon\vecf{\theta}\in\Theta_0$, \citet{KaplanZhuo2021} describe how the type of disagreement depends on the shape of $\Theta_0$, and the specific example of SC is detailed (Section 5.1).
Formally, the frequentist properties are characterized for the Bayesian test that rejects $H_0$ when its posterior probability is below $\alpha$, which has both intuitive and decision-theoretic reasons.

Depending on the ordinal relationship of interest, the Bayesian approach may be more conservative, less conservative, or similar to the frequentist approach.
If only a single inequality is near binding, then the test essentially reduces to a one-dimensional one-sided test, in which case the frequentist and Bayesian assessments generally agree.
If multiple inequalities are involved with $H_0\colon X\SD{1}Y$, then $\Theta_0$ is convex and the frequentist approach favors $H_0$ more than the Bayesian approach, as also pointed out by \citet{Kline2011}.
Conversely, if instead $H_0\colon X\nonSD{1}Y$, then the non-convexity of $\Theta_0$ can make the Bayesian approach favor $H_0$ more.
For $H_0\colon X\SC Y$, $\Theta_0$ is a subset of a half-space, but it has some locally non-convex regions, so the frequentist assessment tends to favor $H_0$ more in small samples, but the Bayesian assessment may favor $H_0$ more in large samples if the true $\vecf{\theta}$ is near a non-convexity of $\Theta_0$; see Section 5.1.2 of \citet{KaplanZhuo2021}.

\section*{Supporting Information}

Additional Supporting Information may be found in the online version of this article at the publisher's website:

~\\
\noindent
Replication files

\end{document}